\documentclass{aa}  
\usepackage{graphicx}
\usepackage{txfonts}
\usepackage{float}
\begin{document} 

   \title{Spatially resolved spectroscopy across stellar surfaces. IV. }

   \subtitle{F, G, \& K-stars: Synthetic 3D spectra at hyper-high resolution }
  
   \author{Dainis Dravins
          \inst{1},
       Hans-G\"{u}nter Ludwig
           \inst{2},
 \and
       Bernd Freytag
          \inst{3} 
}
%
%
\institute{Lund Observatory, Department of Astronomy and Theoretical Physics, Lund University, Box 43, SE-22100 Lund, Sweden\\
              \email{dainis@astro.lu.se}
\and
     Zentrum f\"{u}r Astronomie der Universit\"{a}t Heidelberg, Landessternwarte, K\"{o}nigstuhl, DE--69117 Heidelberg, Germany\\
              \email{hludwig@lsw.uni-heidelberg.de}
\and
     Theoretical Astrophysics, Department of Physics and Astronomy, Uppsala University, Box 516, SE-75120 Uppsala,  Sweden \\
            \email{bernd.freytag@physics.uu.se}
             }
 
\date{Received 26 November, 2020; accepted 18 February, 2021}

 
\abstract
   {High-precision stellar analyses require hydrodynamic 3D modeling.  Such models predict changes across stellar disks of spectral line shapes, asymmetries, and wavelength shifts.  For testing models in stars other than the Sun, spatially resolved observations are feasible from differential spectroscopy during exoplanet transits, retrieving spectra of those stellar surface segments that successively become hidden behind the transiting planet, as demonstrated in Papers~I, II, and III.  }
   {Synthetic high-resolution spectra over extended spectral regions are now available from 3D models.  Similar to other ab initio simulations in astrophysics, these data contain patterns that have not been specifically modeled but may be revealed after analyses to be analogous to those of a large volume of observations. }
   {From five 3D models spanning T$_{\textrm{eff}}$ = 3964--6726\,K (spectral types $\sim$K8\,V--F3\,V), synthetic spectra at hyper-high resolution ($\lambda$/$\Delta\lambda$\,>1,000,000) were analyzed.  Selected \ion{Fe}{i} and \ion{Fe}{ii} lines at various positions across stellar disks were searched for characteristic patterns between different types of lines in the same star and for similar lines between different stars.}
   {Spectral-line patterns are identified for representative photospheric lines of different strengths, excitation potential, and ionization level, thereby encoding the hydrodynamic 3D structure.  Line profiles and bisectors are shown for various stars at different positions across stellar disks.  Absolute convective wavelength shifts are obtained as differences to 1D models, where such shifts do not occur. }
   {Observable relationships for line properties are retrieved from realistically complex synthetic spectra.  Such patterns may also test very detailed 3D modeling, including non-LTE effects.  While present results are obtained at hyper-high spectral resolution, the subsequent Paper~V examines their practical observability at realistically lower resolutions, and in the presence of noise. }
 

\titlerunning{Spatially resolved spectroscopy across stellar surfaces. IV.}
\authorrunning{D. Dravins, H.-G.Ludwig, \& B.Freytag}
\maketitle

\section{Introduction}

This project concerns the atmospheric structure and detailed spectra of solar-type stars, and detection schemes in searches for Earth-like exoplanets in orbits around them.  We aim to record high-resolution spectra across spatially resolved stellar surfaces and to identify signatures from surface features such as granulation or magnetically active regions.  In Papers~I, II, and III of this series \citep{dravinsetal17a, dravinsetal17b, dravinsetal18}, a method for such spatially resolved spectroscopy was elaborated and applied to observations of the G0~V star HD\,209458 and the K1~V star HD\,189733A (``Alopex'').  Differential spectroscopy during exoplanet transits retrieves spectra from those stellar surface portions that temporarily become hidden during successive transit epochs.  Tests of 3D hydrodynamic models of photospheric convection become possible through comparisons to synthetic spectral-line profiles at various center-to-limb positions, computed as temporal and spatial averages over their simulation sequences.  This Paper~IV explores observable spectral-line shapes and shifts, in the range 400-700 nm, from complete synthetic spectra of different spectral types obtained from hydrodynamic 3D simulations. The subsequent Paper~V \citep{dravinsetal21} will examine the practical observability of specific spectral features (and their variability in time) when observed with realistic instrumentation.

This paper is organized as follows: Section 2 describes the role of ab-initio simulations and the challenges in finding Earth-like exoplanets.  Section 3 recalls some general aspects of 3D modeling of stellar atmospheres and describes the presently used sequence of CO\,$^5$BOLD models and the synthetic spectra obtained from those.  Section 4 explains the role of corresponding 1D models.  Section 5 treats the selection of representative spectral lines, Section 6 examines how the intensity profiles and bisectors in different classes of lines differ among various stellar models and how they change across stellar disks, while Section 7 concludes with a discussion of issues still outstanding for the future. 

\section{Spectra from 3D hydrodynamic atmospheres}

High-precision determinations of stellar properties require a detailed modeling of stellar surfaces and an understanding of how the emerging spectrum of radiation is formed.  The structure of convective motions can be modeled in some detail with 3D and time-dependent hydrodynamics; such model photospheres are now established as the most realistic descriptions of stellar outer layers, at least in solar-type stars \citep{beecketal13a, freytagetal12, kupkamuthsam17, leenaarts20, leitneretal17, magicetal13, nordlundetal09,  trampedachetal13, tremblayetal13}, with a general agreement among different model families \citep{beecketal12, pereiraetal13}. 

For 3D solar models, it has been possible to observationally verify the predicted features of granulation and associated atmospheric structures against high-resolution images and spectra across the solar photosphere.  At lower spatial resolution, when averaged over many granular structures, spectral line profiles exhibit gradual changes between the solar disk center and the limb \citep{koesterkeetal08}; there is somewhat different behavior between various atomic species for lines of different strengths, excitation potentials, and ionization levels \citep[e.g.,][]{balthasar88, lindetal17, ramellietal17, takedaueno19}.  Most lines show a general increase of equivalent width toward the limb; some lines that have particularly high excitation potentials instead show a decrease.  On the Sun, the convective blueshift normally decreases from disk center toward the limb but amplitudes and line asymmetries depend on line strength.

Such hydrodynamic modeling is computationally demanding and, for solar-type stars, is currently feasible only over simulation volumes that comprise a tiny fraction of the star, and necessarily involves certain physical, mathematical, and numerical approximations.  For example, effects of oscillations have to be approximated and larger-scale phenomena such as supergranulation or meridional flows are not well handled since they do not fit inside the simulation volumes.  Once the atmospheric structures are modeled from such 3D simulations, the ensuing spectral synthesis involves some further simplifications.  These could include treating only single and idealized spectral lines, approximating opacities over broad spectral regions, assuming local thermodynamic equilibrium (LTE), using only 1D radiative transfer, or some other simplification.   

Testing and verifying such 3D models for other stars remains rather challenging because of the difficulty in spatially resolving their disks.  In this project, a method for spatially resolved spectroscopy across stellar surfaces was elaborated and applied to observations of a first few stars.  Paper~I \citep{dravinsetal17a} demonstrated how differential spectroscopy during exoplanet transits can recover spectra from those stellar surface portions that temporarily become hidden during successive transit epochs.  A Jupiter-sized exoplanet, covering $\sim$1$\%$ of the disk of a solar-type star, provides a fairly good spatial resolution along the transit path of the planet.  This size of exoplanet also gives adequate averaging over surface inhomogeneities since it covers $\sim$10,000 granules out of the total $\sim$10$^{6}$ present on solar-type stars.  The transit path is rather precisely determined from the Rossiter-McLaughlin effect, pinpointing the location of samplings across the stellar surface.  Such observationally retrieved spectral lines are free from stellar rotational broadening and can be confronted with synthetic spectral lines computed from 3D models.

Features identified in the G0~V star HD\,209458 \citep[Paper~II,][]{dravinsetal17b} include the gradual increase of photospheric \ion{Fe}{i} spectral-line broadening from stellar disk center toward the limb.  Such behavior agrees with 3D predictions for such and hotter stars, where horizontal velocities in granulation are greater than vertical velocities, and are well visible in the line-forming layers near the limb, producing an increased Doppler broadening there.  

For the cooler K1~V star HD\,189733A (Alopex), such an effect is neither expected nor observed \citep[Paper~III;][]{dravinsetal18}.  Further modeling constraints may come from the strongest (partly chromospheric) lines, such as \ion{Ca}{ii} H and K, or \ion{Na}{i} D$_1$ and D$_2$, where limb brightening in their wings relative to the adjacent continuum has been seen \citep{czeslaetal15}.  However, such lines may be further affected by stellar magnetic activity, similar to the case for H$\alpha$ \citep{cauleyetal17}. 

Such differential spectroscopy is challenging in terms of the photometric precision needed; sensible reconstructions require high-resolution spectra with signal-to-noise ratios $\gtrsim$\,5,000.  While such values are not (yet) reachable in single exposures of individual spectral lines, they may be realized for the line-rich spectra of cooler stars by averaging over numerous lines with similar parameters.  Many Jupiter-size transiting planets will likely be discovered from ongoing and planned exoplanet searches, and the brightest host stars of these planets will be suitable targets for spatially resolved spectroscopy.  These are also likely to be high-priority targets for exoplanet studies, assuring their observation with high-resolution instruments on large telescopes.  However, it is not obvious precisely which characteristic signatures can realistically be retrieved for different types of stars and which should best be confronted with model simulations.  Anticipating such observations, this Paper~IV surveys complete synthetic spectra for stars of different temperatures to identify patterns that are unique for 3D models and plausible to identify in realistically complex spectra with a multitude of often overlapping or blending lines.  The following Paper~V \citep{dravinsetal21} evaluates which spectral features should be practically observable, depending on the signal-to-noise and spectral resolution that can be attained.

Other 3D effects might be detectable with optical interferometry, where closure phases could serve as an indicator for stellar surface inhomogeneities \citep{chiavassaetal12, chiavassaetal14}.  Also, the short-term photometric variability during a planetary transit carries some imprints from the amplitudes and spatial scales of brightness fluctuations across the stellar surface \citep{chiavassaetal17, morrisetal20a, sarkaretal18, sulisetal20}.

\subsection{The quest for Earth-like exoplanets}

An outstanding challenge is to find near Earth-like exoplanets, whose sizes, masses, and orbits are comparable to the terrestrial values.  One plausible path toward their detection is to identify the wobble in radial velocity caused by the stellar motion around the barycenter of the system, induced by the orbital motion of the planet(s).  Although the tiny amplitude induced by the Earth on the Sun amounts to at most only 0.1 m\,s$^{-1}$,  instrumental sensitivities are now beginning to reach such levels \citep{halletal18}; the main limitations are no longer technical, but lie in understanding the complexities of atmospheric dynamics and spectral line formation, manifest as a jittering of the apparent radial velocity and as a flickering in photometric brightness \citep{cegla19, fischeretal16}. 

Given that the radial-velocity wobble induced by a small exoplanet is much smaller than stellar microvariability, there must be a way to calibrate and correct for the effects of the latter.  Several authors have studied empirical correlations between various stellar activity indices and corresponding excursions in apparent radial velocity.  Although revealing the effects of magnetic and other processes and permitting some mitigation of stellar velocity signals, these correlations do not reach sufficient predictability to also permit the identification of low-mass exoplanets.  We doubt that such types of empirical correlations will be able to approach the required precisions and therefore different and spectroscopically more targeted approaches will be needed.  

Spatially resolved spectra can verify and constrain 3D simulations and these models can be used to predict and calibrate temporal variability in the emergent spectrum of the host star.  Fluctuations in a hydrodynamic atmosphere affect many parameters in concert.  Once such processes are understood from more basic principles, it should be possible to identify which combinations of spectral parameters best serve as proxies for jittering in radial velocity.  In particular, for quantities that can be measured from the ground, it should be possible to estimate the observational precision needed, enabling the push toward sub-m\,s$^{-1}$ precisions required for exoEarth detections.  An understanding of the impact from magnetic regions will also be required, but spatially resolved spectra can also be retrieved from active-region granulation plages and starspots whenever a planet happens to pass in front of them. These issues will be discussed further in Paper~V.  

The method demands very high spectral resolution, very stable wavelength calibration, and exceptionally low photometric noise.  Although not commonplace in the past, such demands are met by numerous new spectrometers designed to enable radial-velocity searches for exoplanets.  Three-dimensional signatures across the disk of HD209458 could already be clearly identified from data at $\lambda$/$\Delta\lambda$ $\sim$80,000 in Paper II, in which exposures covered only half a transit across the stellar disk.  Data from new instruments, such as ESPRESSO at the ESO VLT\footnote{ESPRESSO, the Echelle SPectrograph for Rocky Exoplanets and Stable Spectroscopic Observations, at VLT, the Very Large Telescope of ESO, the European Southern Observatory}, empower more precise studies.  For example, a recent exoplanet study used that spectrometer to observe the same star with twice higher spectral resolution, a much improved wavelength scale, and with more numerous exposures of comparable photometric precision during two full transits \citep{casasayasetal21}.  Without awaiting improved performance with future extremely large telescopes, suitable new targets will likely be found by ongoing exoplanet surveys.  These surveys keep finding host stars spanning broad temperature ranges and the brightest of those, which also have some large transiting planets, will be primary candidates for future studies. 

A further motivation to study spatially resolved stellar spectra is that atmospheric properties of any transiting exoplanet have to be deduced from subtle differences against the background stellar spectrum, which thus must be precisely known.  This requires a knowledge of the varying stellar line profiles at the positions along the transit path of the planet \citep{rackhametal18, yanetal17}; those are the features that are filtered through the exoplanetary atmosphere, not the spectrum of the flux from the full stellar disk.  Additionally, the spectrum of the stellar flux that remains during exoplanetary transit (then displaying the Rossiter-McLaughlin effect) depends not only on stellar rotation but is also modified by how line profiles change across the stellar disk.

\begin{figure*}
\sidecaption
\centering
\includegraphics[width=18cm]{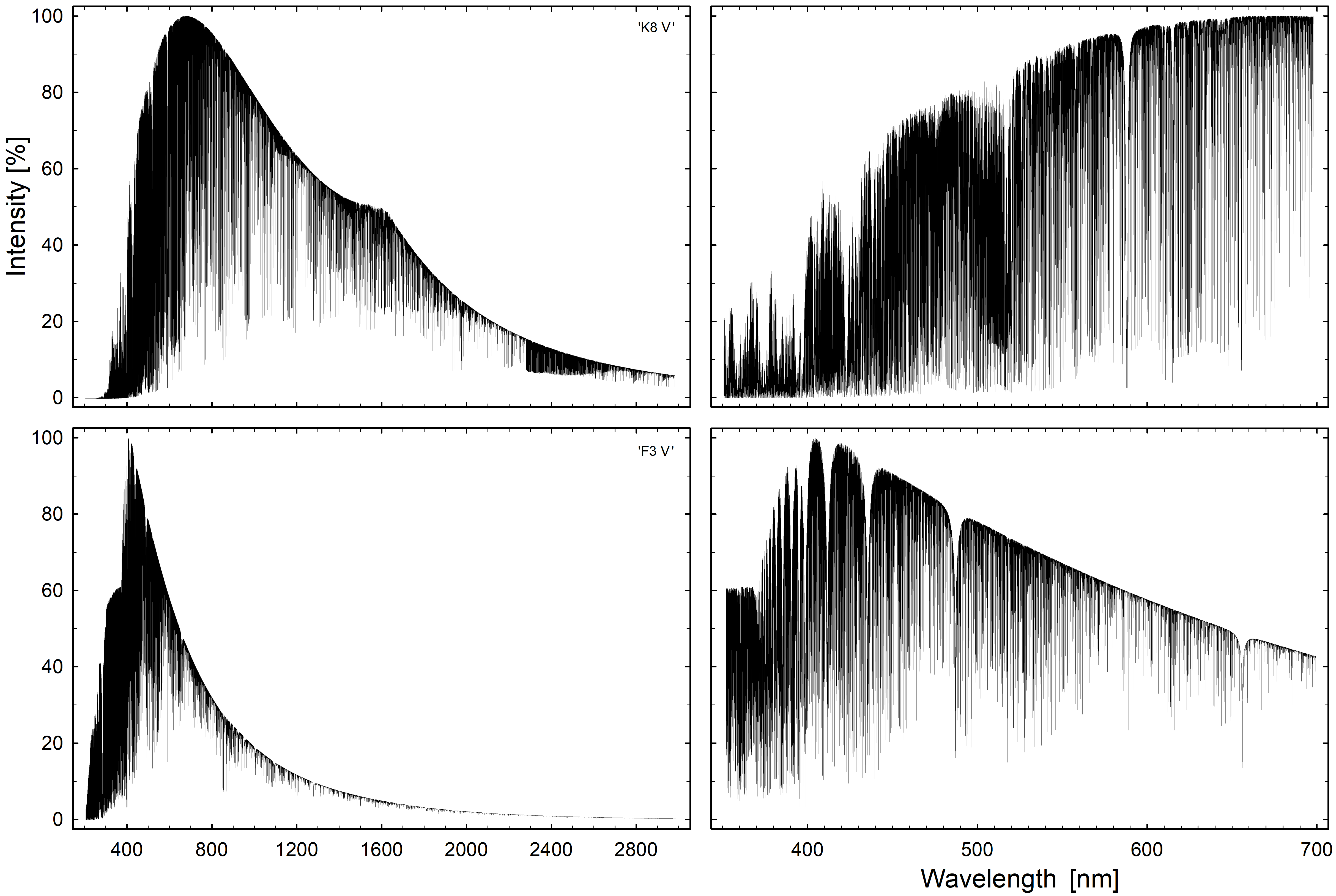}
\caption{Synthetic full-disk flux over the full spectral range for the coolest and hottest models in the current sample, normalized to the peak of the flux.  Top row: T$_{\textrm{eff}}$ =  3964K, `K8~V'; bottom: T$_{\textrm{eff}}$ = 6726 K, `F3~V'.    Here and elsewhere, stellar rotation is assumed to be zero. }
\label{fig:full_spectra}
\end{figure*}

\subsection{The new role of ab-initio spectral simulations}

The role of simulations goes beyond merely reproducing observations.  Such ab-initio computations are based upon fundamental physical principles and do not contain any freely adjustable parameters that could be tweaked to force a better agreement with observational data.  Of course, parameters used for the input are chosen to correspond to the best understanding of, for example, atmospheric opacities, radiative transfer, and boundary conditions, while model extents and stepsizes are constrained by the available computational capacity.  The output from such simulations is somewhat unpredictable in that the simulations in principle should reproduce the natural processes, some outcomes of which may not be known nor expected.  Similar to analogous work in other fields, such as large-scale cosmological simulations, it should be rewarding to explore the resulting data volumes in searches for new types of patterns or relations between parameters that previously have not been recognized but in retrospect might be understood after examining the simulation sequences.

Recently, a major step has been possible in 3D stellar simulations: the computation of complete stellar spectra \citep{chiavassaetal18, ludwigetal21}.  In previous works, for example, in Papers~I, II, and III and \citet{beecketal13b}, profiles of single idealized and isolated spectral lines were computed using the simulated atmospheric structure as a grid of space- and time-varying model atmospheres; these profiles were then compared to observations.  Although such synthetic line profiles may well demonstrate the types of effects present, such idealized cases only offer incomplete representations of actual stellar spectra when facing observational realities.  This current enhancement involves complete and complex stellar spectra, incorporating practically all relevant spectral lines from databases such as VALD\footnote{VALD, The Vienna Atomic Line Database of atomic and molecular transition parameters of astronomical interest} and others \citep{heiteretal15, ryabchikovaetal15}, and doing so at an exceptionally high spectral resolution over all practically observable spectral ranges.  The details in resulting spectra depend on a complex interplay between, for example, blending of multiple spectral lines, temperature-dependent atmospheric opacity in different wavelength regions or different atmospheric inhomogeneities, or as seen under different viewing angles across the stellar disks.  In this Paper~IV, we begin to explore spectral features resulting from such simulations.  These originate from a sequence of CO\,$^5$BOLD model atmospheres for solar-metallicity dwarf stars with T$_{\textrm{eff}}$ between 3964\,K and 6726\,K, corresponding to approximate spectral types K8\,V--F3\,V.  Synthetic spectra were computed with sampling stepsizes $\Delta\lambda$ corresponding to hyper-high spectral resolutions ($\lambda$/$\Delta\lambda>$1,000,000)\footnote{The term hyper-high is used to denote spectral resolutions $\lambda$/$\Delta\lambda\gtrsim$\,10$^6$ since ultra-high is already in common use for describing spectrometers with the much lower resolutions of only $\sim$200,000.}; some 3 million spectral data points span the ultraviolet to the infrared, 200 nm to 3,000 nm, for each of 21 angular directions of emerging radiation and for each of some 20 instances in time during the simulation sequence \citep{ludwigetal21}.  Those instances were selected to be sufficiently apart in time to ensure largely uncorrelated flow patterns between successive samples.  Hyper-high resolution (with comparable data previously encountered only in solar spectroscopy) is necessary to fully resolve intrinsic line asymmetries and to obtain wavelength shifts on the level of m\,s$^{-1}$.  To segregate those asymmetries from those arising owing to blends, and to obtain absolute wavelength shifts irrespective of errors in laboratory wavelengths, these 3D spectra are matched against similar hyper-high resolution spectra computed from static 1D models.  There, convective line shifts do not occur, not even in principle, and unblended lines appear symmetric at their nominal laboratory wavelength positions: the differences to 3D profiles thus isolate the specifics arising in the inhomogeneous and dynamic photospheres. 

The analysis of such spectra represents a new class of stellar studies, which is somewhat analogous to what already has been possible for simulated atmospheric structures and in numerical simulations within other fields.  The simulated spectra have characteristics analogous to a large volume of observational data, which have to be categorized, analyzed, and interpreted.  Even if still modest by the standards of some other large-scale simulations \citep[e.g.,][]{nelsonetal15}, the data volume is significant by the standards of spectra of individual stars.  Any detailed analyses of the many thousands of spectral lines in the data would be rather challenging, and in this work we can only consider some limited aspects of the data.  These synthetic spectra are surveyed for apparently unblended lines with different strengths, excitation potentials, and ionization levels, each of which contribute characteristic signatures of line shapes, asymmetries and wavelength shifts.  In the following Paper~V, these hyper-high resolution data will be degraded to more ordinary spectrometer values to appreciate what signatures then are still preserved and may realistically be observed.  Also, correlations in the time variability between different spectral parameters will be examined with the aim to identify possible proxies for the jittering in apparent radial velocity, which might be used to adjust the latter to true stellar motion, as required for the detection of small exoplanets.

\section{CO\,$^5$BOLD hydrodynamic models}

CO\,$^5$BOLD 3D model simulations are used, comprising small but statistically representative box-in-a-star volumes in stellar surface layers.  With an original $xyz$ spatial resolution of 140$\times$140$\times$150 points, both optically thin and thick regions are included \citep{freytagetal12}.  In our current model selection, the range of their averaged temperatures spans T$_{\textrm{eff}}$ between 3964\,K and 6726\,K and, for convenience, these models are referred to with their approximate corresponding spectral types as `K8~V', `K2~V', `G1~V', `F7~V' and `F3~V' (Table \ref{table:synthetic_spectra}).  Metallicities are solar throughout, [Fe/H] = 0, with the opacity handling described by \citet{ludwigsteffen13}.  

Spectral synthesis \citep{ludwigetal21} was performed assuming LTE.  Synthetic spectral lines were computed from these simulation volumes, sampling every third point in each horizontal direction, resulting in spatial grids of 47$\times$47 locations for each of about 20 instances sampled during the simulation.  For each of these, the flux was computed for 21 emergent angles.  There are four angles of $\theta$, the inclination against the normal to the stellar surface, corresponding to the center-to-limb positions of $\mu$ = cos\,$\theta$ = 1, 0.79, 0.41, and 0.09.  For the off-center positions $\mu$ = 0.79 and 0.41, emergent spectra were computed for eight different azimuth angles $\psi$, every 45$^{\circ}$ along the circle, while the $\mu$ = 0.09 position close to the limb was computed for every 90$^{\circ}$.  It must be noted, however, that the spectrum synthesis very close to the limb has additional uncertainties since these 3D models are not optimal for the more spherical curvature there.

\begin{figure}
\centering
\includegraphics[width=\hsize]{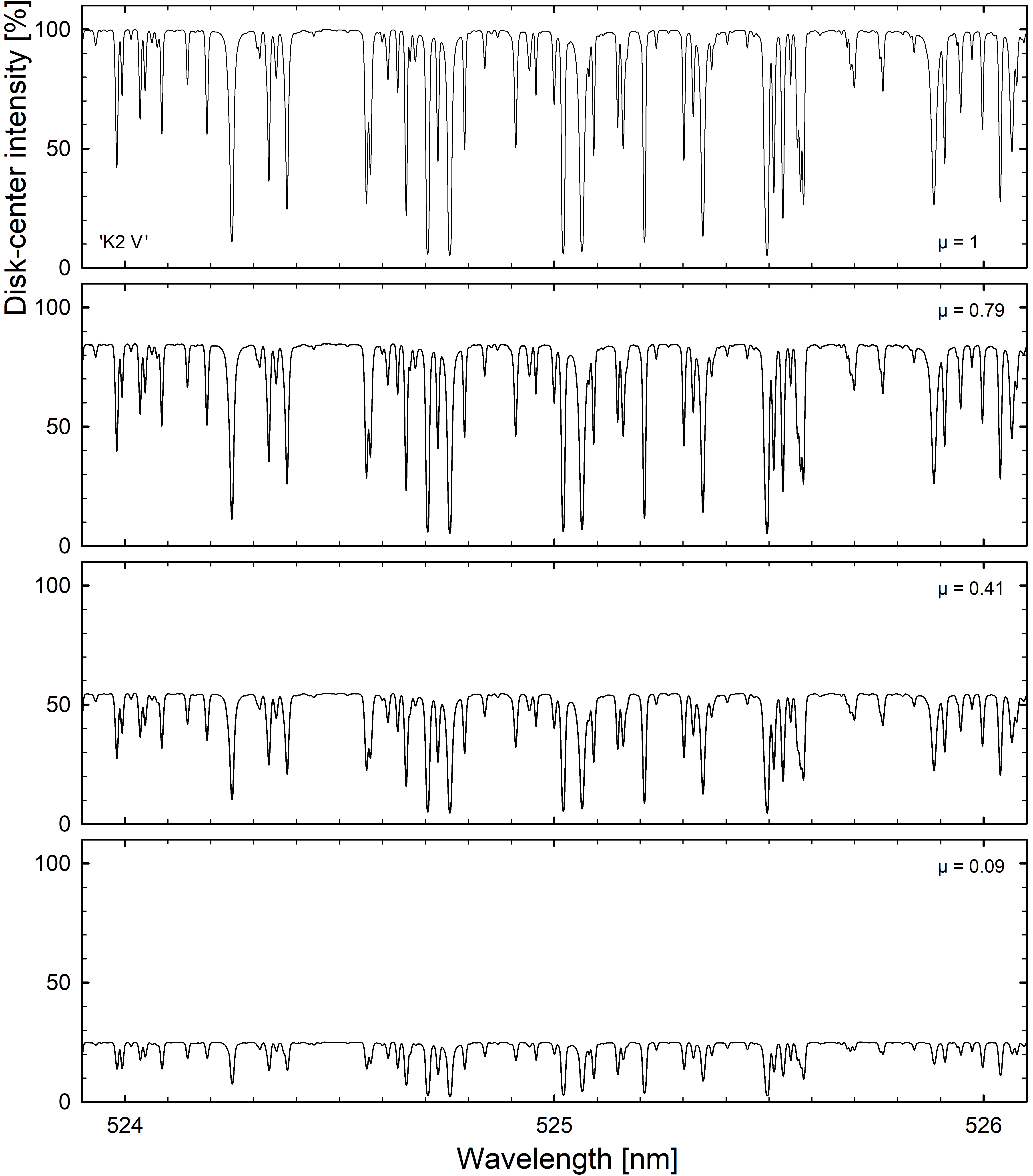}
\caption{Synthetic spectra over a small spectral range in the model for T$_{\textrm{eff}}$ = 4982 K (`K2~V'), at positions from stellar disk center (top frame, $\mu$ = 1) to near the limb (bottom, $\mu$ = 0.09).  The intensity scale is that of stellar disk center; changing continuum levels reflect limb darkening at these particular wavelengths.}
\label{fig:spectral_portions}
\end{figure}

\begin{figure}
\centering
\includegraphics[width=\hsize]{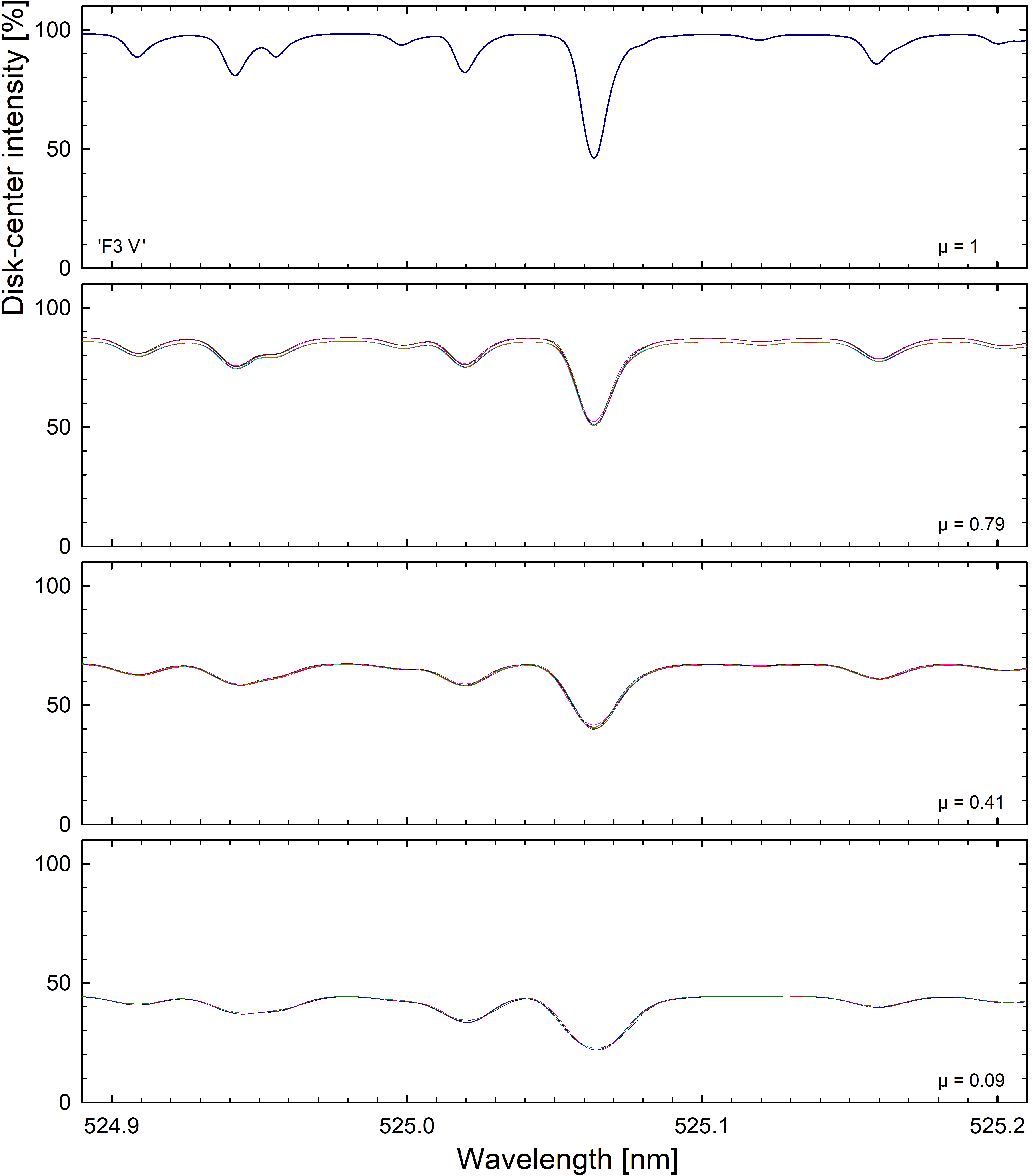}
\caption{Synthetic spectra in a narrow spectral segment, from stellar disk center (top frame, $\mu$ = 1) to near the limb for the `F3~V' model at T$_{\textrm{eff}}$ =  6726 K.  For off-center positions, spectra from different azimuth angles are plotted, illustrating the variation across steps of 45$^{\circ}$ or 90$^{\circ}$ in azimuth. To see those subtle differences, this figure should be viewed highly magnified. }
\label{fig:spectral_details}
\end{figure}

\begin{table}
\caption{Synthetic spectra computed from CO\,$^5$BOLD atmospheres.  Models in this sequence for main-sequence stars are referred to with their approximate spectral types.  Data points making up the synthetic spectra were sampled in steps of uniform values of $\lambda$/$\Delta\lambda$.  Further details are in Table \ref{table:co5bold_models}.  }             
\label{table:synthetic_spectra}      
\centering          
\begin{tabular}{c c c c}
\hline     
\noalign{\smallskip}
Spectral type & T$_\textrm{eff}$ [K] &  Data points & $\lambda$/$\Delta\lambda$ \\ 
\noalign{\smallskip}
`K8~V'  &  3964  &  3,357,216  &  1,240,000 \\
`K2~V'  &  4982  &  3,357,192  &  1,238,000 \\
`G1~V'  &  5865  &  3,020,160  &  1,114,000 \\
`F7~V'  &  6233  &  2,901,480  & 1,071,000 \\
`F3~V'  &  6726  &  2,726,184  &  1,007,000 \\
\hline   
\end{tabular}
\end{table}

For each model, synthetic spectra were computed for the full wavelength range from 200 nm in the ultraviolet to 3 $\mu$m in the infrared, sampled with a constant stepsize in velocity units of m\,s$^{-1}$, and thus the spectral resolution in terms of $\lambda$/$\Delta\lambda$ is constant throughout.  It differs slightly among the models since the stepsize was chosen to equal 30\% of the width of an iron line at the lowest temperature encountered within each respective simulation volume; for the cooler models this is somewhat smaller, and the resolution thus somewhat greater.  The resulting $\lambda$/$\Delta\lambda$ values range between 1,007,000 and 1,240,000 for the different models, with between 2.7 and 3.4 million data points making up each spectrum.  Overall model properties are in Tables \ref{table:synthetic_spectra} and \ref{table:co5bold_models}; for computational details, see \citet{ludwigetal21}.

Fig.\ \ref{fig:full_spectra} shows examples of synthetic spectra over the full spectral range for the coolest and hottest models in the current group.  On these highly compressed plots, well visible features for `K8~V'  include a hump around $\lambda$~1600 nm caused by opacities from the negative hydrogen ion, and strong molecular bands further in the infrared.  The `F3~V' model has its flux shifted toward shorter wavelengths, and displays strong hydrogen Balmer lines with a pronounced Balmer jump decrement. 

While Fig.\ \ref{fig:full_spectra} shows spectra from the full stellar disks, Fig.\ \ref{fig:spectral_portions} gives an example over a 2 nm interval from the `K2~V' model of spatially resolved spectra at different center-to-limb positions.  A third example illustrating the character of the simulated spectra is in Fig.\ \ref{fig:spectral_details}.  This very narrow segment in wavelength from the `F3~V' model illustrates not only its center-to-limb variations, but also the amplitude of the emergent spectrum in 21 different angular directions.  At disk center ($\mu$ = 1), there is only one unique direction toward the observer but at $\mu$ = 0.79, 0.41, and 0.09, emergent spectra are shown for different azimuthal angles. The successively more detailed views depicted by Figs.\ \ref{fig:full_spectra} to \ref{fig:spectral_details} thus illustrate the substantial information content of such spectra.

\section{Corresponding 1D models}

For each 3D model, a corresponding 1D version was also computed based on ATLAS model atmospheres \citep{ludwigetal21}.  These models were tailored to have the same effective temperature, surface gravity and chemical abundance as the CO\,$^5$BOLD variants.  Since spectral line synthesis was made using practically the same opacities and radiative transfer algorithms as in 3D, this permits direct comparisons between these classes of models.  Adjustable 1D quantities include the mixing-length parameter (set to 1.25) and the microturbulence, which was computed for several different values, whereafter some macroturbulent line broadening could be applied.  However, we have no ambition to optimally fit any 1D model to either observations or 3D variants; their purpose here is only to provide a reference point in deducing absolute wavelength shifts of spectral lines. 

In dynamic and inhomogeneous 3D photospheres, convective wavelength shifts occur because the greater photon contributions from hot, bright, and rising elements (thus locally blueshifted) normally dominate over those from cool, dark, and sinking gases, most often causing a convective blueshift.  While this effect can be clearly isolated in idealized models of single spectral lines (as in Papers~I, II, and III), the situation is less straightforward in more realistic simulations with numerous overlapping and blending lines.  Spectral lines may become distorted, asymmetric, and shifted in wavelength as a result of even very faint overlapping or blending lines.  However, since the same input line lists are used for computing 1D and 3D models, the same line blends appear in both models.  Since 1D models have no atmospheric dynamics to cause asymmetries, a truly unblended line must appear fully symmetric, with a vertical bisector located at its laboratory wavelength position (neglecting gravitational redshift), and a comparison to 3D reveals the line asymmetry and the absolute value of its wavelength shift.  Since the wavelength scales are common for the various center-to-limb positions, such a calibration also provides the absolute center-to-limb variations of the convective line shifts.  This method is also immune to plausible errors in laboratory wavelengths.  Such errors are known to correspond to several tens of m\,s$^{-1}$ for strong lines of low excitation-potential but often increase to perhaps $\sim$100 m\,s$^{-1}$ for weaker high-excitation lines or such from ionized species.  Of course, different errors in laboratory wavelengths of closely adjacent lines also enter into these simulated spectra and may cause spurious line asymmetries.  However, reasonable wavelength errors for such blending lines are small compared to their widths and, by far, weaker blending lines do not induce spurious line asymmetries comparable to those induced by convective motions.

\begin{figure}
\centering
\includegraphics[width=\hsize]{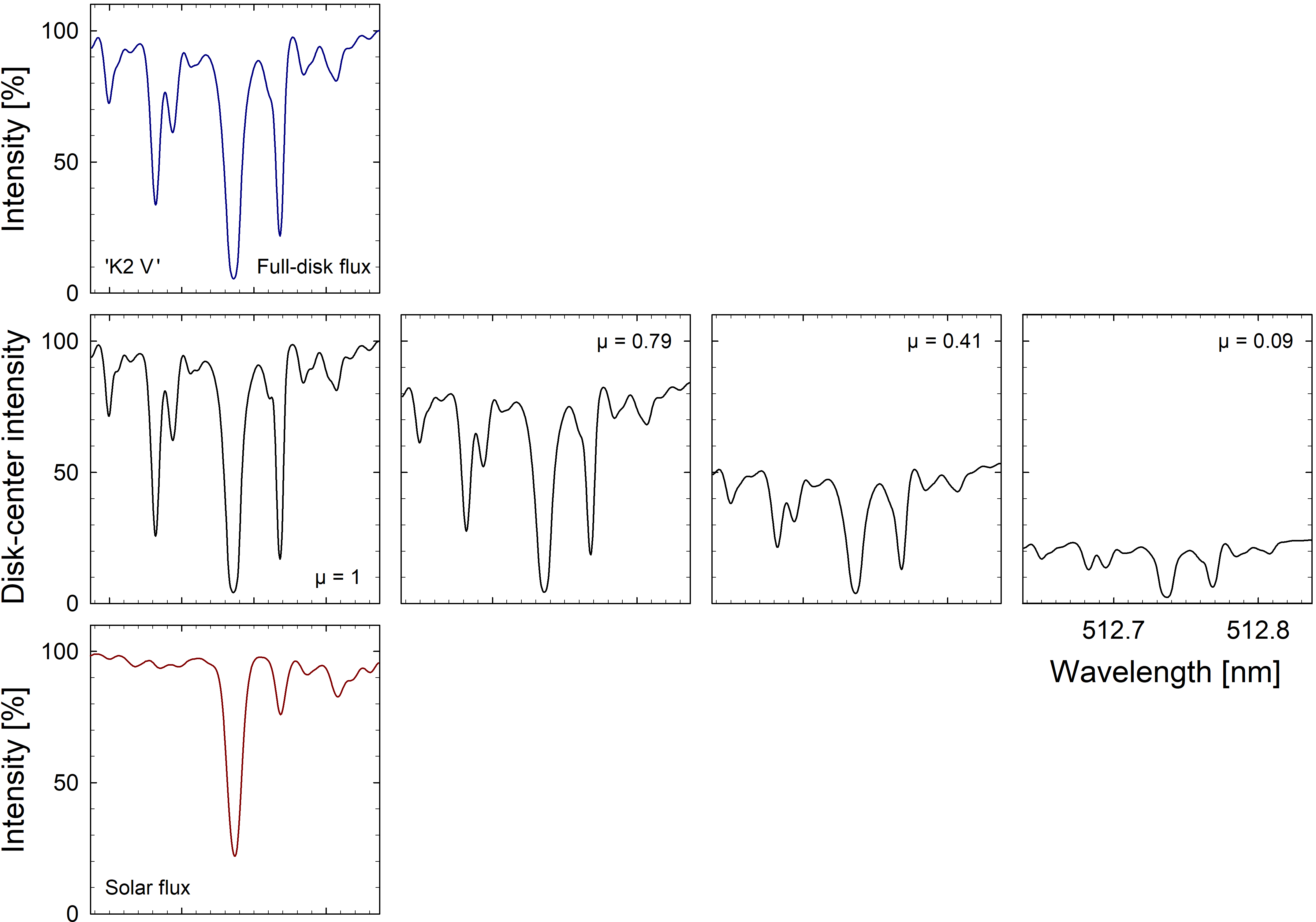}
\caption{Example of a selected line from the `K2~V' model (very strong \ion{Fe}{i}, low excitation potential, $\lambda$~512.737 nm).  Besides the changing profiles at different stellar disk positions, the top frame shows the synthetic full-disk flux, and the bottom frame -- as a reference -- the same line in a high-resolution solar flux spectrum atlas \citep{kuruczetal84}. }  
\label{fig:line_selection}
\end{figure}

\section{\ion{Fe}{i} and \ion{Fe}{ii} spectral line selection}

We wish to study the behavior of different classes of realistically measurable spectral lines across stellar disks to see how they encode atmospheric properties and how their signatures vary among stars of different spectral types.  The number of measurable spectral lines is huge and is almost overwhelming; already very narrow wavelength ranges, such as the 2 nm sample in Fig.~2, include a wealth of lines, hinting at the challenges in selecting the most suitable lines.  It would take a disproportionate effort to try to identify larger fractions of them to search for unblended candidates common to multiple models.  However, the temperature range of our models brackets the solar value, and lines that can be followed at temperatures both cooler and hotter than solar must reasonably be essentially unblended in the Sun as well.  Similar to previous studies, lines from iron are preferred thanks to both their multitude and the lack of significant atomic structure complications (large mass, thus small thermal broadening); one isotope dominates, which is even-even in its proton-neutron numbers, thus there is no nuclear spin and no hyperfine splitting.  

Therefore, in starting the selection process, we used listings of unblended solar Fe lines, as originally obtained from high-fidelity solar spectrum atlases, in particular the listings of \ion{Fe}{i} lines by \citet{stenflolindegren77} and \ion{Fe}{ii} data cited in \citet{dravins08}.  We grouped these solar lines according to line-strength, excitation potential, ionization level, and wavelength region, and within each such grouping, we selected sets of clean and similar lines as representative of that grouping.  An example of one such line is in Fig.\ \ref{fig:line_selection}.  The profiles for the off-center positions $\mu$ = 0.79, 0.41 and 0.09 are now taken as averages over all azimuth angles.  That line widths tend to increase toward the limb is already seen in that the profile for the integrated full-disk spectrum is slightly broader than that at disk center; the stellar rotation is zero, thus there is no rotational broadening.  Figure \ref{fig:line_selection} also shows its corresponding profile in integrated solar flux at a resolution $\lambda$/$\Delta\lambda$ $\sim$500,000 from the atlas by \citet{kuruczetal84}, which was used in the line selection process.

\begin{figure*}
\centering
\includegraphics[width=18cm]{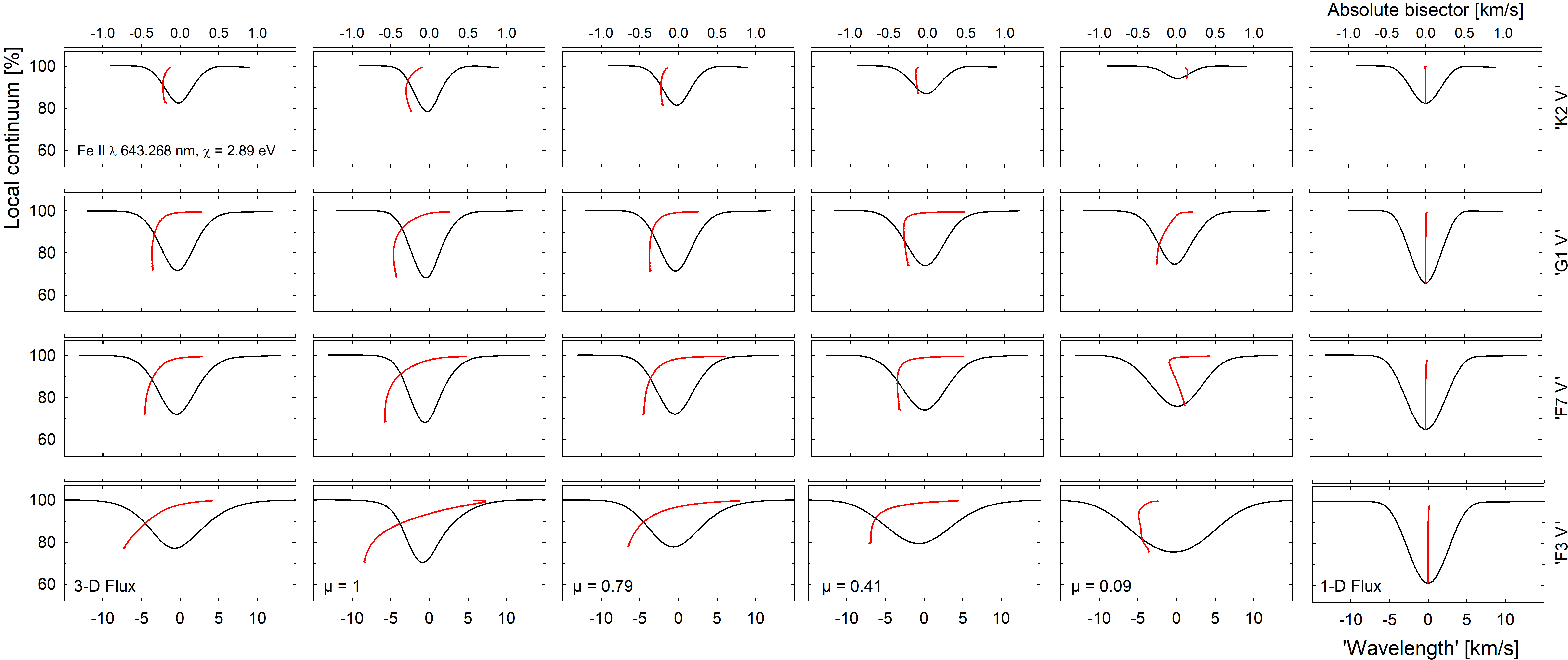}
\caption{Line profiles and bisectors on an absolute wavelength scale from 3D and 1D models, exemplified by the medium-strong \ion{Fe}{ii} 643.26831 nm line in the `K2~V', `G1~V', `F7~V', and `F3~V' models (in rows from top down).  Leftmost column shows spectra from the full stellar disks, followed by columns for $\mu$ = 1, 0.79, 0.41, and 0.09.  In the rightmost column, the vertical bisectors for the 1D profiles confirm that the line is unblended, enabling placement of the 3D bisectors on an absolute wavelength scale (now converted to velocity units).  The red bisector curves are expanded tenfold relative to the intensity profiles; their axes are at the top.} 
\label{fig:line_sample}
\end{figure*}

Six line groups were selected for each of \ion{Fe}{i} and \ion{Fe}{ii}, with four lines in each group.  These thus represent 12 different classes of \ion{Fe}{i} and \ion{Fe}{ii} lines in strength and excitation potential but of course do not cover all possible subsets.  Even if these lines were selected to permit tracking the line from one stellar model to next (Fig.\ \ref{fig:line_sample}), spectral features change considerably with changing temperature (Fig.\ \ref{fig:full_spectra}) and not many lines remain distinct and wholly unblended between several models and/or across all disk positions. 

Given the richness of cool-star spectra, we imagine further subdividing with respect to wavelength region, Land{\'e} $\varg$$_{\textrm{eff}}$-factor (indicating sensitivity to magnetic fields), and separately for different atomic species other than \ion{Fe}.  However, with finer subdivisions, the quality criteria for spectral line inclusion must necessarily decrease since there simply does not exist enough clean lines to represent all possible parameter combinations.  Although idealized synthetic lines can be computed for all sorts of parameters, the finite diversity of real spectral lines constitutes a limit to the information content of stellar spectra \citep{dravins08}. The selected line groups are listed in Table \ref{table:table_fe_lines}.  For \ion{Fe}{i}, the Land{\'e} $\varg$$_{\textrm{eff}}$-factors are also listed.  As discussed in Paper~V, it is foreseen that forthcoming work may include simulated spectra from stellar magnetic regions as well, where this line selection could prove useful.

\section{Line profiles and absolute bisector shifts}

To enable direct comparisons between lines, the wavelength scales were transformed from wavelengths in air to velocity units of the corresponding Doppler velocity relative to the respective line center.  For the absolute wavelength scale, that is, wavelength positions relative to a system at rest with respect to the star (neglecting gravitational redshifts), the reference point is obtained from the 1D models.  Those models produce no asymmetries from atmospheric dynamics but most lines are still somewhat asymmetric; because of the presence of overlapping and blending lines, the same blends contribute similar asymmetries in 3D models.  When measured under finite spectrometer resolution, the amount of blending also depends on whether the instrumental profile touches nearby lines, which can be more noticeable than in our clean high-fidelity spectra.  Therefore, before selecting any candidate line, we verified that the 1D profiles also remained essentially unblended with the spectral resolution downgraded to $\lambda$/$\Delta\lambda$ = 75,000, well within the range of current spectrometers.  Fig.\ \ref{fig:line_sample} shows an example of one specific spectral line in 1 among the 12 line-groups.  Very noticeable changes in line shapes and asymmetries are seen when going from temperatures of $\sim$5000 to 6700\,K.  Before examining the consistency of such behavior among different lines, the changes across stellar disks and between models at different temperatures, we first place the specific challenges in context.

\subsection{True stellar radial motion}

Understanding line asymmetries and wavelength shifts is coupled to the need to precisely determine true stellar center-of-mass motions from spectroscopy.  As noted above, a plausible scheme for detecting low-mass planets in Earth-like orbits around solar-like stars is to identify the tiny wobble in radial velocity of the host star in its reflex motion around the common barycenter with the orbiting planet.  However, the miniscule signal from such a planet remains buried in the much greater fluctuations in wavelength arising in the dynamic stellar atmosphere, which thus must be somehow calibrated and corrected for.  Besides temporal variability, issues arise because stellar spectral lines are asymmetric and their wavelengths and the deduced radial velocity are not precisely defined quantities since they depend on exactly which portions of the spectral lines are measured, and – as discussed in Paper V – also vary with instrumental resolution.  Further, irrespective of planet searches, precision studies of motions between components of binary stars or those in open clusters require an understanding of the relations between spectroscopically measured wavelength displacements and physical stellar motion.  

To obtain absolute values for radial motion, the gravitational redshifts affecting stellar spectra also need to be understood.  These change only slightly along the main sequence but major differences exist between dwarf and giant stars.  Some information on convective line shifts may then be obtained by comparing dwarfs and giants in open clusters such as M67 \citep{pasquinietal11} or between components of binary stars in well-determined orbits such as $\alpha$\,Cen~A and B \citep{moschellaetal21, pourbaixetal02}.  It may also be noted that the gravitational redshift need not be a constant for any one star since its precise value depends on the formation height of any particular spectral line \citep{lindegrendravins03} and on possible stellar oscillations that could affect the stellar diameter.  Already a very small change in stellar radius could mimic the signal from an Earth-like planet \citep{ceglaetal12}.  

Theoretically deducing true radial motion from the apparent velocity thus requires spectral-line modeling that realistically reproduces the line asymmetries induced by convection, oscillations, and perhaps by further effects.  Furthermore, a value for the gravitational redshift is required.  Such calculations for the Sun have been made by \citet{cruzrodriguezetal11}, \citet{gonzalezhernandezetal20}, \citet{gonzalezmanriqueetal20} and others.   For specific spectral lines, complications with additional wavelength displacements (even neglecting magnetic effects) may arise owing to isotopic shifts \citep[e.g.,][]{kramida20, leenaartsetal14} or atomic hyperfine structure \citep[e.g.,][]{lefebvreetal03, scottetal15}.  

For other stars, synthetic line profiles including convective line shifts for full-disk spectra were computed by \citet{allendeetal13, dravinsnordlund90b} and \citet{magicetal14}.  Observational searches for different line shifts among various spectral types and the dependence on their activity levels include \citet{meunieretal17a, meunieretal17b}.  Signatures vary among different spectral lines and depend on line strength, excitation potential, and wavelength region.  For example, the lower opacity in solar infrared lines carries signatures from deeper atmospheric volumes than lines in the visual \citep{milicetal19}.

In principle, the problem might perhaps be circumvented by instead determining radial motions from second-order effects in astrometry.  Although this is doable for stars in nearby clusters such as the Hyades (whose stars share a common space motion), and where required accuracies are available from space astrometry, such methods do not deliver short-term velocity variations \citep{debruijneetal01, dravinsetal99, leaoetal19, madsenetal02}.

\subsection{The Sun seen as a star}

A primary reference is the spectrum of the Sun, which allows us to recognize types of features expected in stellar spectra.  However, even the highest-fidelity solar atlases display subtle discrepancies due to differences in precisely how the data were recorded.  This is not only a consequence of instrumental specifics, but also exactly where on the Sun the spectra were sampled and when: at what phase in the sunspot activity cycle, at which time of year (viewing the solar rotational axis under what angle), with what size and shape of the entrance aperture (how much averaging of solar rotation), or for how long a time period (how much line broadening due to oscillation averaging).  Differences between solar disk-center atlases \citep{doerretal16} indicate the level of consistency that is present or that may be expected.  It may be sobering to recall that -- as has been underscored by \citet{kurucz09} and others -- that there does not yet exist any high-fidelity solar or stellar optical spectrum recorded from outside the atmosphere.  All ground-based spectra are contaminated by superposed telluric lines, the details depending on the humidity and pressure of the local air masses \citep{cunhaetal14, kauschetal15, smetteetal15, xuesongetal19}.

The spectrum of the Sun seen as a star has been obtained with various methods for full-disk integration.  The G\"{o}ttingen atlas of integrated solar flux in the optical was recorded with a hyper-high resolution ($\sim$10$^{6}$) Fourier transform spectrometer.  This instrument has a wavelength scale believed to be accurate to $\sim$\,10\,m\,s$^{-1}$, enabling critical comparisons between different solar atlases regarding convective line shifts \citep{lemkereiners16, reinersetal16}.  Among stellar spectrometers, HARPS\footnote{High Accuracy Radial velocity Planet Searcher at the La Silla 3.6~m telescope of ESO, the European Southern Observatory} has been used to record atlases of reflected sunlight from Ceres and Ganymede \citep{molarocenturion11, molaromonai12} as well as for spectra of the Moon \citep{molaroetal13}.  A solar-disk integration telescope feeding PEPSI\footnote{Potsdam Echelle Polarimetric and Spectroscopic Instrument at LBT, the Large Binocular Telescope} recorded the flux spectrum at a resolution of $\sim$270,000 \citep{strassmeieretal18}.  

High-resolution studies of individual spectral lines in solar quiet regions reveal how spectral lines are sculpted by convection.  Using a laser frequency comb for absolute wavelength calibration in the LARS spectrometer on the VTT\footnote{Laser Absolute Reference Spectrograph on the Vacuum Tower Telescope at the Teide Observatory on Tenerife} solar telescope, \citet{lohnerbottcheretal18, lohnerbottcheretal19} and \citet{stiefetal19} carried out detailed studies of various lines in the visible.  In particular, their data demonstrate well how the line asymmetry (bisector shape) degrades very significantly when going from hyper-high spectral resolution to more ordinary values.  For disk-center profiles, the span between the line bottom and the maximum excursion of the bisector toward shorter wavelengths gradually shrinks from typically $\sim$200 m\,s$^{-1}$ to just $\sim$30 m\,s$^{-1}$ when decreasing the resolution $\lambda$/$\Delta\lambda$ from 700,000 to 100,000.  The amounts differ among lines of different strength and depth; examples are provided in \citet[][their figure 12]{lohnerbottcheretal18}; \citet[][their figure 5]{lohnerbottcheretal19}, or \citet[][their figure 5]{stiefetal19}.   

From solar flux atlases covering wide spectral regions, line asymmetries and their bisector spread can be statistically assessed, showing both similarities and differences \citep{dravins08, grayoostra18}.  However, bisector shapes depend significantly on the spectral resolution realized, and most bisectors are further distorted owing to blends with overlapping or adjacent lines.  Also, if one particular value is given for a wavelength position of a line, that depends on exactly how wavelength is defined in the fitting to the line profile.  Absolute wavelength shifts are normally measured relative to the corresponding laboratory wavelengths which often have errors in the range 10-100\,m\,s$^{-1}$, normally limited by physical conditions in the laboratory light sources such as their pressure or internal electric and magnetic fields.  Much of the spread in published plots of solar line shifts is certainly due to such spread in laboratory wavelengths rather than the measuring imprecision in solar spectra \citep[e.g.,][]{gonzalezhernandezetal20}.

Given this background, in this Paper~IV we first explore spectral-line signatures in full hyper-high resolution to understand what is present in stellar spectra before degrading the synthetic data to observationally realistic resolutions in Paper~V.  Also, given the complexities of real spectra with their multitude of overlapping and blending lines, we select real spectral features to provide examples of what might be observable rather than in only idealized and isolated lines.

\begin{figure}[!]
\centering
\includegraphics[width=\hsize]{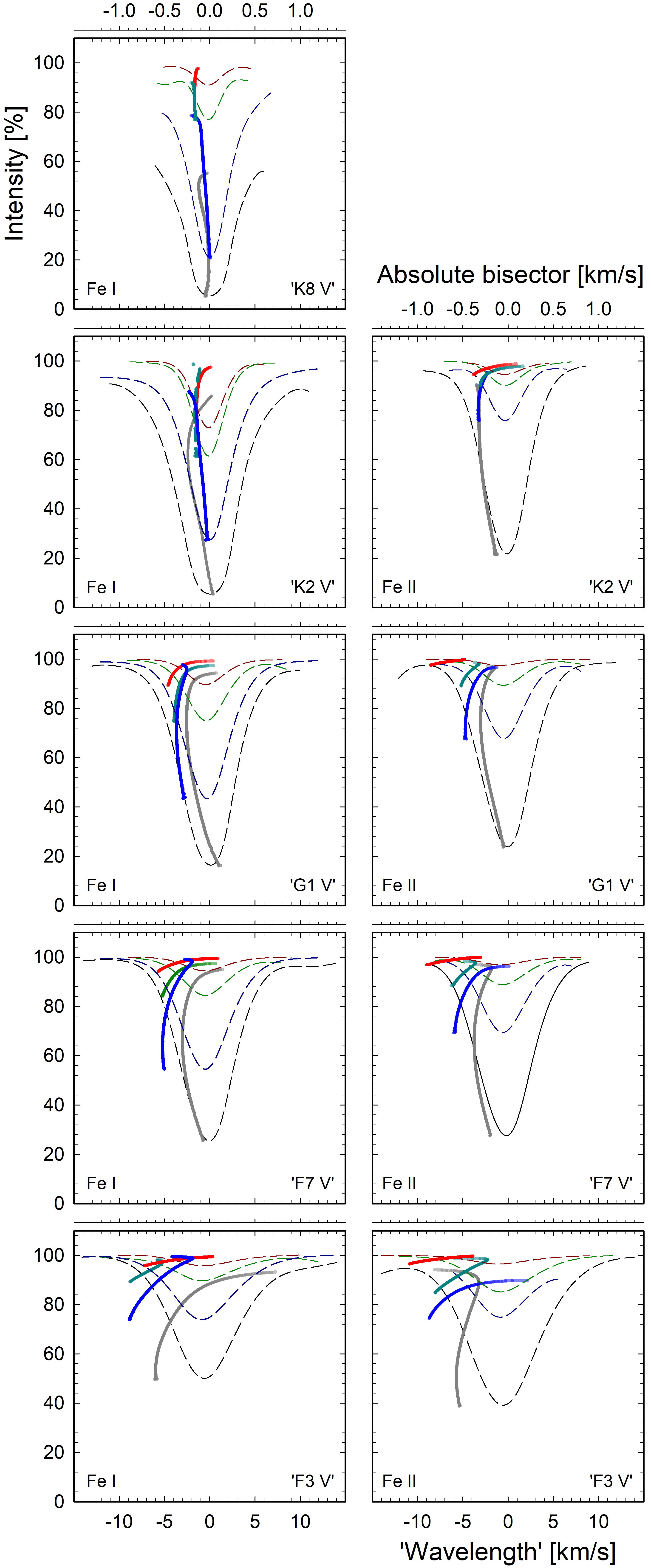}
\caption{ 
 Bisectors (tenfold expanded top axes) and profiles (bottom) for differently strong \ion{Fe}{i} (left column) and \ion{Fe}{ii} lines (right) in the flux from the full stellar disks for five different stellar models.  Temperatures increase top down for the models `K8~V’, `K2~V’, `G1~V’, `F7~V’, and `F3~V’. Stellar rotation is zero, gravitational redshifts are neglected. For the coolest model, only lines from \ion{Fe}{i} were well measurable. Continuum intensity scale is that of the overall stellar spectrum; in crowded spectra of the coolest models, not all lines reach the full continuum.  }
\label{fig:line_similarities}
\end{figure}

\begin{figure}[h]
\centering
\includegraphics[width=\hsize]{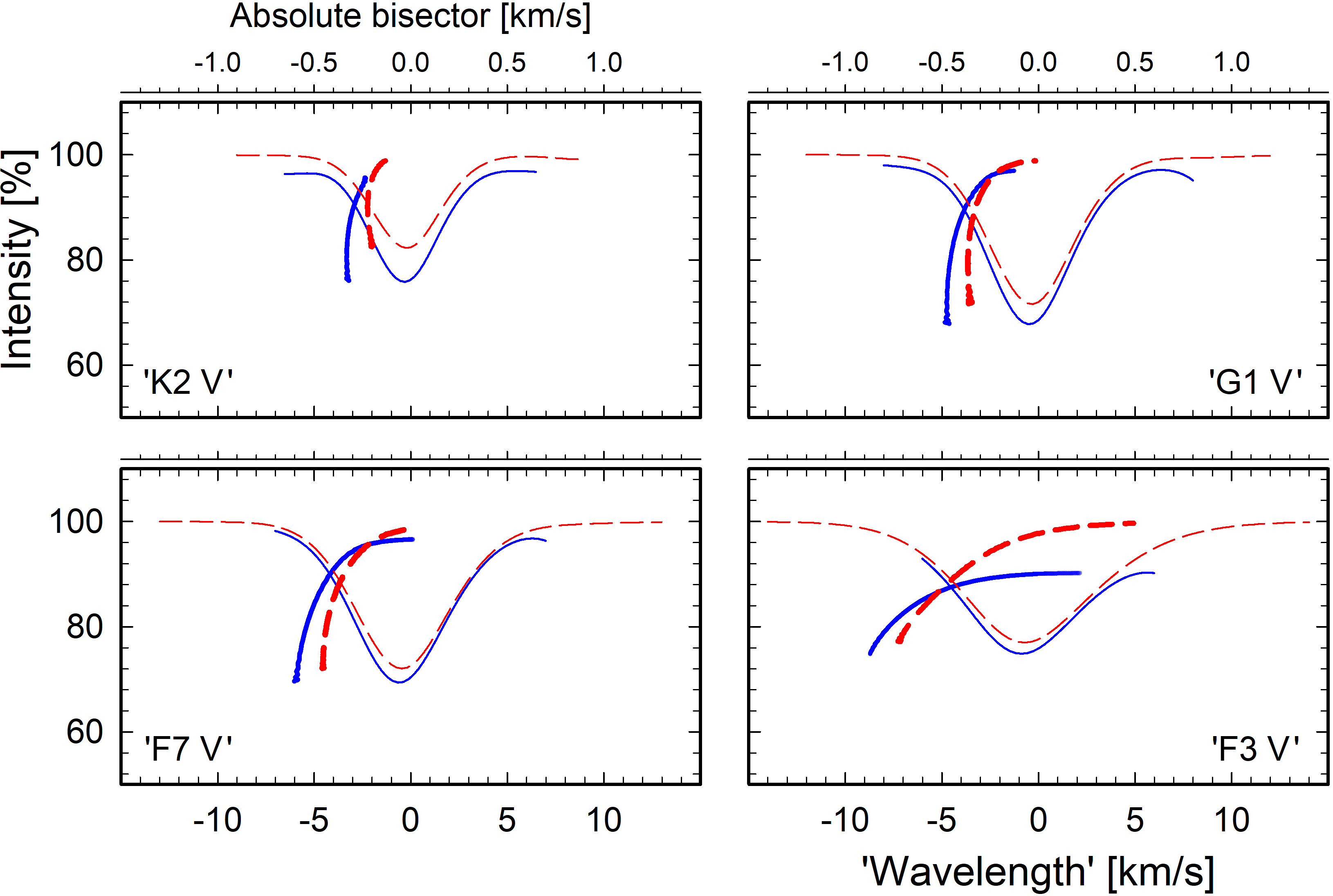}
\caption{Convective line shifts depend on wavelength region.  Bisectors (tenfold expanded top axes) and profiles (bottom) for spectral lines of similar strength and excitation potential in short- and long-wavelength regions show effects of greater granulation contrast in the blue than in the red.  Solid blue curves: \ion{Fe}{ii} $\lambda$ 467.017 nm, $\chi$ = 2.58 eV; dashed red: \ion{Fe}{ii} $\lambda$ 643.268 nm, $\chi$ = 2.89 eV. }
\label{fig:color_dependence}
\end{figure}

\subsection{Line asymmetries in integrated starlight }

 Before examining variations across stellar disks, we first illustrate spectral line shapes and asymmetries in the flux from full stellar disks that result from the simulations.  This permits us to compare the present work with previous modeling and observations.  Data in Figures \ref{fig:line_similarities} and \ref{fig:color_dependence} are at the full hyper-high spectral resolution for nonrotating stars.  Comparisons with observations need to consider that lower spectrometer resolutions weaken line asymmetries, while averaged wavelength positions are less distorted.  For example, effects at the moderate spectral resolutions on the Gaia spacecraft were examined by \citet{allendeetal13}.  Line asymmetries are further modified by stellar rotational broadening, possibly in a nontrivial manner, depending on the viewing angle relative to the stellar axis of rotation and the center-to-limb line changes of the line profile \citep{dravinsnordlund90b, graytoner85, smithetal87}.   

The most pronounced variation in spectral line asymmetries is between lines of different strengths.  For Figure~\ref{fig:line_similarities}, lines were selected within the \ion{Fe}{i} and \ion{Fe}{ii} groups in Table~\ref{table:table_fe_lines}, creating a representative sequence of differently deep line profiles and bisectors for each model.  Given the calibration against 1D models, the bisector wavelengths are absolute and permit us to examine the changes between differently strong lines in each star, the gradual alteration of the bisector patterns with stellar temperature, and the distinction between neutral and ionized lines. 

The `G1~V’ model parameters lie very close to solar values and the bisector patterns are well recognized from previous solar studies; the convective blueshifts of around 300 m\,s$^{-1}$ are gradually increasing for weaker \ion{Fe}{i} lines.  For \ion{Fe}{ii}, asymmetries and blueshifts are somewhat enhanced.  For other spectral types, however, previous work is more limited. 

With $\lambda$/$\Delta\lambda$ $\sim$100,000, \citet{gray82, gray09} measured bisector wavelength spans (denoting them as the `‘third signature of granulation’’) for G- and K-type stars of different luminosity classes, noting that they follow the solar pattern but with amplitudes increasing with stellar temperature and decreasing surface gravity.  Also \citet{allendeetal02a, allendeetal02b} found bisector patterns qualitatively similar to solar in various (mostly evolved) cool stars, although a more detailed analysis of Procyon (F5~IV-V) showed rather more pronounced convective signatures.

Using a great number of HARPS spectra from the ESO archive, \citet{meunieretal17a, meunieretal17b} examined main-sequence stars between K4 and F7.   Wavelength references were obtained from fitting line bottoms, with the changing line shift as a function of line depth indicated by the bisector slope.  These authors found this slope to depend on stellar temperature and, if extrapolated, approaching a vertical bisector for T$_{\textrm{eff}}$ = 4680~K.  Such a value is bracketed by our two coolest models in Fig.~\ref{fig:line_similarities}, at 3964 and 4982 K, where the bisector curvature has almost disappeared.

Theoretical bisector patterns from STAGGER grid models for temperatures 4500--6500~K show how convective signatures increase for higher temperatures and for weaker lines, also revealing ionization-level and excitation-potential dependences \citep{magicetal14}.  A few earlier calculations were by \citet{dravinsnordlund90b}. These previous studies appear consistent with the bisector patterns in Fig.~\ref{fig:line_similarities}, even if there are slight model differences and observations do not yet cover all types of spectral lines in all types of stars.

\begin{figure*}
\sidecaption
 \centering
 \includegraphics[width=12.9cm]{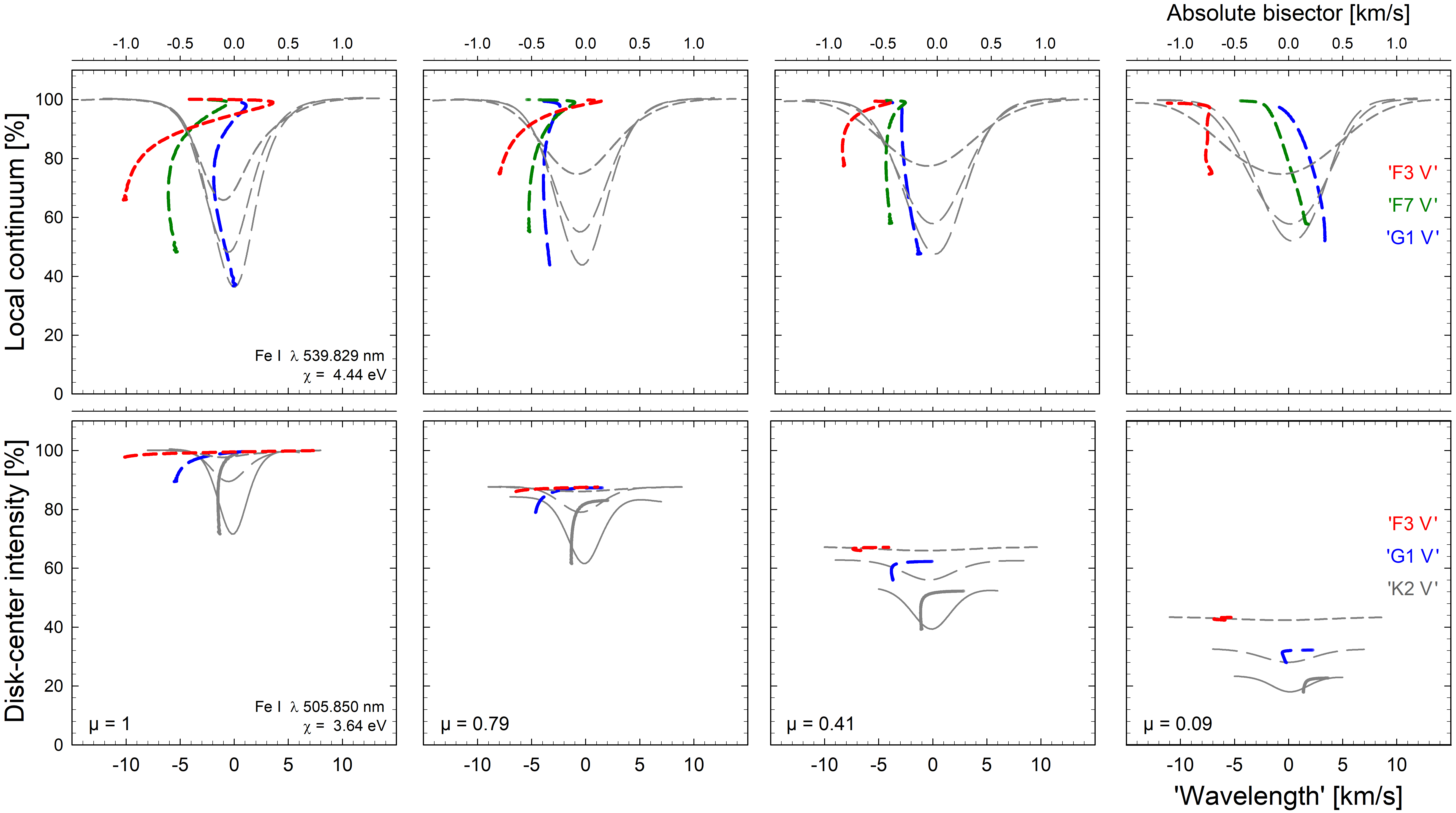}
     \caption{Changing \ion{Fe}{i} profile shapes (axes at bottom) and bisectors (tenfold expanded axes at top of each frame) across different stars.  The top row shows a strong line in 3D models for `G1~V' (long-dashed), `F7~V' (medium-dashed), and `F3~V' (short-dashed); the bottom row shows a weak line for 'K2 V' (solid), 'G1 V', and 'F3 V'.  For the top row, the local spectral continua are normalized to 100\% at each center-to-limb position while the intensity scales at bottom are those of the stellar disk centers, showing how the limb darkening changes with stellar temperature.  Bisectors (top axes) are absolute in wavelength and unaffected by blending lines; in corresponding 1D models, all bisectors appear as vertical lines at practically zero displacement. } 
     \label{fig:changing_shapes}
\end{figure*}

\subsection{Difference between wavelength regions}

Besides the primary dependence on line strength, lesser dependences exist between ionization levels (as in Fig.~\ref{fig:line_similarities}), excitation potentials (as discussed below), and on wavelength regions (illustrated in Fig.\ref{fig:color_dependence}).   In solar data, this dependence on wavelength region is seen observationally and theoretically \citep[e.g.,][and references therein]{asplundetal00, hamiltonlester99}: convective blueshifts are enhanced at shorter optical wavelengths.  This is expected, since a temperature difference between thermally radiating hotter and cooler elements causes a greater intensity contrast at shorter wavelengths, enhancing the statistical dominance of blueshifted line components from the hotter and brighter granules.  For other stars, the same trend was noted by \citet{meunieretal17a} in their study of HARPS spectra.

For Fig.~\ref{fig:color_dependence}, clean \ion{Fe}{ii} lines of closely similar strength and with closely similar excitation potential were selected from different wavelength regions ($\lambda$ 467.017 nm, $\chi$ = 2.58 eV and $\lambda$ 643.268 nm, $\chi$ = 2.89 eV).  The long-wavelength lines are very marginally weaker than the short-wavelength lines; however, if they follow the same trends as in Fig.~\ref{fig:line_similarities},we would expect marginally greater blueshifts for the weaker lines, while the opposite is observed.  These spectra are for the flux from the full stellar disks; the center-to-limb behavior of the 643.268 nm line is represented in Fig.~\ref{fig:line_sample}

\begin{figure*}
 \includegraphics[width=18cm]{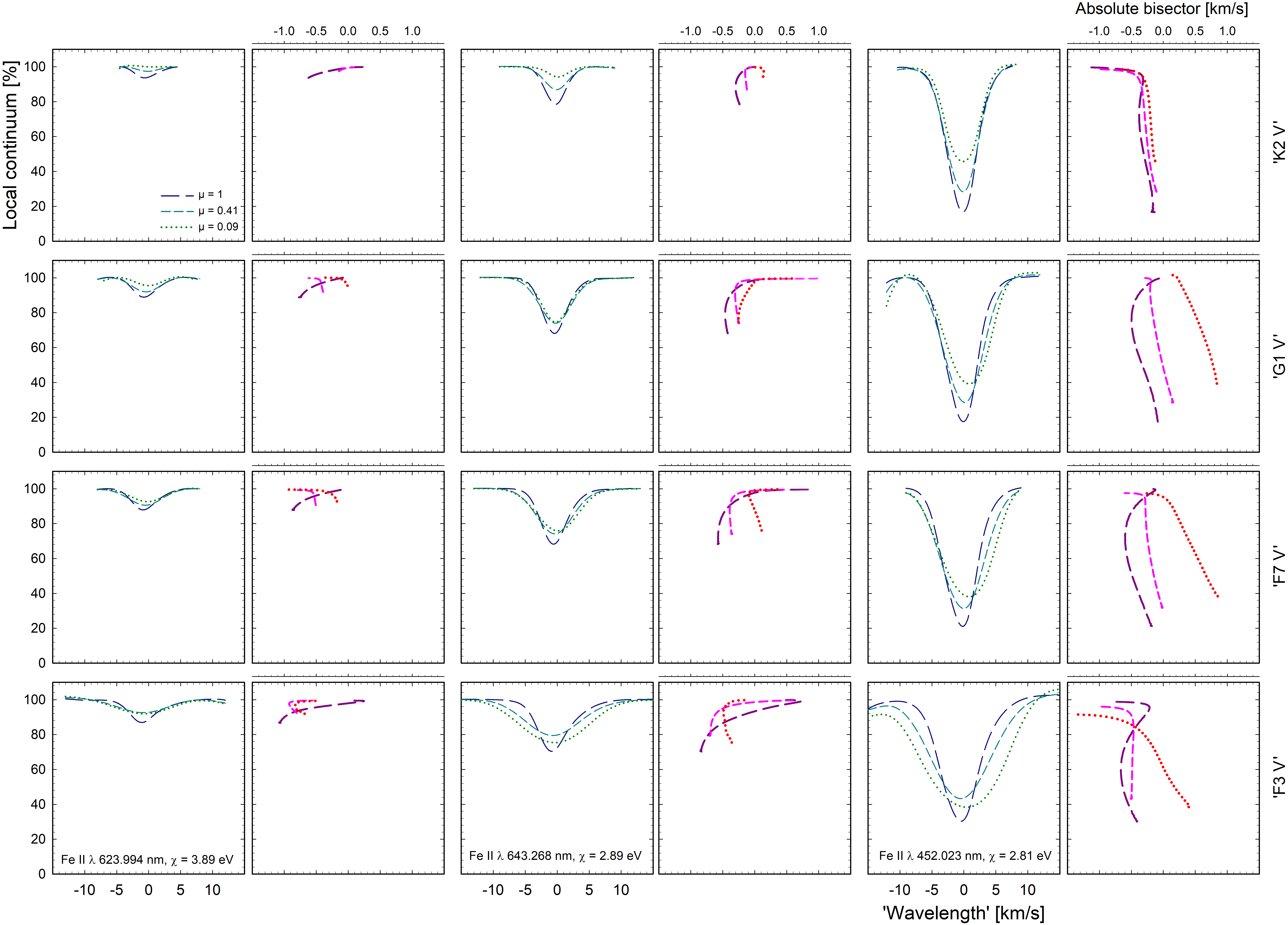}
     \caption{Line-strength dependence for \ion{Fe}{ii} lines.  The profiles (axes at bottom of each frame) and bisectors (tenfold expanded axes at top) are followed through four different models of successively increasing temperature.   The two leftmost columns show a weak line; two central columns one of medium strength, with a strong line in the two rightmost columns.  From top down: `K2~V', `G1~V', `F7~V', and `F3~V'.  The long-dashed profiles and bisectors are for disk centers at {$\mu$} = 1, short-dashed for {$\mu$} = 0.41, and dotted for near-limb data at {$\mu$} = 0.09. } 
     \label{fig:line_strength}
\end{figure*}

\subsection{Similar lines across different stars}

Fig.\ \ref{fig:changing_shapes} shows representative changes of \ion{Fe}{i} line profile shapes and bisectors across different stars.  In general, line widths increase toward the limb, where the generally greater amplitude of horizontal motions in stellar granulation contribute more Doppler broadening along the line of sight.  Rising gases from below decelerate and turn over at heights comparable to the vertical pressure scale height, which is smaller than the horizontal extent of granules; thus horizontal velocities must be greater to satisfy mass conservation.  The changing patterns seen in Fig.\ \ref{fig:changing_shapes} can be thus understood from such modeling details. 

On solar-like G-type stars, the full amplitudes of granule brightness and velocities are slightly veiled beneath the line-forming layers.  This causes the brightness contrast to decrease with height, manifest in effects such as a decreased granulation contrast in strong lines formed at higher atmospheric levels, a distinct dependence of convective blueshift on line strength and also its decrease from disk center toward the limb (where it may even overshoot into a redshift).  In cooler K-type stars, the granulation contrast is substantially lower since the structures are more strongly veiled by atmospheric opacity.  The fully developed granulation is not strictly a surface phenomenon in this case, but rather relates to those somewhat deeper layers where most energy flux converts from convective to radiative.  Although this largely hidden granulation is not deep down in terms of linear depth -- maybe only some 100 km -- the corresponding optical depths are substantial for spectral line formation, and emerging line profiles are expected to carry but modest signatures of the processes in the deeper layers \citep{chiavassaetal18, dravinsnordlund90b, magicetal13, nordlunddravins90, ramirezetal09, tremblayetal13}.  

By contrast, hotter F-type stars display essentially naked granulation with temperature and velocity contrasts fully visible at the optically thin surface.  The differing atmospheric opacities in regions of different temperature contribute to make the surface corrugated (in the optical depth $\tau$ = 1 sense), making the 3D granular structure visible from its side near the limb as well \citep{allendeetal02b, chiavassaetal12, chiavassaetal18, dravinsnordlund90a, dravinsnordlund90b}.  Such a geometry also exposes the full amplitude of the horizontal velocity field to an outside observer, causing a correspondingly greater line broadening.  The increasingly vigorous convection in hotter stars is also reflected in greater convective blueshifts at disk center while the corrugated surfaces may cause the already large convective blueshift to further increase from disk center toward the limb (contrary to a simple picture with only radial motions on a smooth star).  Near the limb, we preferentially observe the near sides of the sloping and rolling granular hills facing the observer, where gas flows are directed mainly toward the observer, thus contributing blueshifted line components.  The equivalent redshifted components, however, remain invisible behind these hills.  In some of the strongest lines, an analogous phenomenon is also seen in somewhat cooler stars, where the greater opacities apparently contribute to shaping corrugated stellar surfaces at small continuum optical depths (Fig.\ \ref{fig:line_strength}).  The details are different between weak and strong lines, lines of various excitation potentials, and those of separate ionization levels.  If corresponding observations could be secured, they should convey very detailed information about conditions in stellar photospheres.

\subsection{Different lines across the same star}

The most pronounced differences between spectral lines are among those of different strength.  Weak and strong lines preferentially form in volumes at different atmospheric heights and since granular convection decays rapidly with height around solar temperatures, this results in significant differences in line asymmetries, shifts, widths and their respective center-to-limb behavior.   Fig. \ref{fig:line_strength} illustrates the changing shapes and shifts of weak, medium, and strong \ion{Fe}{ii} lines across stellar disks.  In this case, the same lines remained unblended and could be followed through four stellar models from `K2~V' to `F3~V'.  To limit cluttering the plot, {$\mu$} = 0.79 profiles are omitted but those do not differ much from the disk-center profiles.  For the weaker lines, convective blueshifts increase with stellar temperature without drastic change in bisector shapes.  However, the strongest line in the hottest models has different center-to-limb runs, and shows a backward bend near the continuum (the so-called  ``blueward hook''). 

Such a feature was first seen observationally in the strongest lines of Procyon (F5~IV-V) when observed under spectral resolutions ~$\sim$200,000, and later given a theoretical explanation by \citet{allendeetal02b}.  This is expected from a superposition of differently shaped line components from granules and from intergranular lanes.  The steeper temperature gradients in the rising granules produce strong, blueshifted, and saturated line components with extended Lorentzian wings.  Their contribution to the blue flank of the spatially averaged line also affects the intensity in the red flank, while the usually redshifted components from intergranular lanes are more Gaussian-like, having an absorption that only extends over a shorter wavelength interval.  Close to the continuum, a small blueshifted intensity depression in the outer line wing may already dominate the bisector, producing the blueward hook.

\begin{figure*}
\sidecaption
 \includegraphics[width=12.9cm]{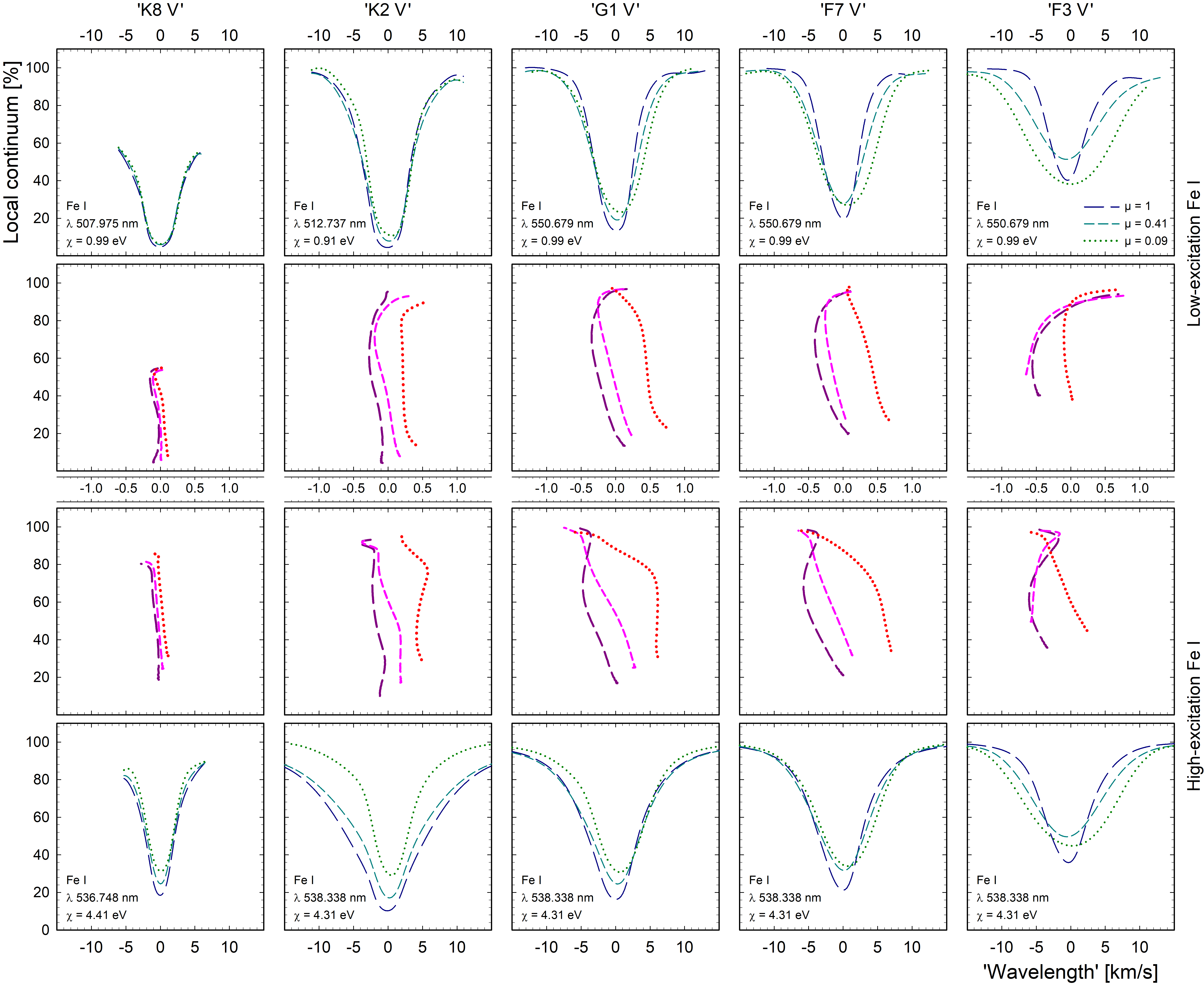}
     \caption{Dependence on excitation potential for \ion{Fe}{i} lines of similar strength in five different models.  Left to right: `K8~V'. `K2~V', `G1~V', `F7~V', and `F3~V'.  Low-excitation profiles and bisectors are indicated in the top rows, high-excitation ones at bottom.  Long-dashed profiles and bisectors are shown for stellar disk-center positions $\mu$ = 1; short-dashed for $\mu$ = 0.41, and dotted for $\mu$ = 0.09.  Bisector scales (at center) are expanded tenfold relative to the line profiles.  } 
     \label{fig:excitation_potential}
\end{figure*}

\subsubsection{Non-LTE line formation}

Such spectral-line signatures with a blueward hook appear promising for the identification of departures from LTE, not only for stars as a whole, but separately within various atmospheric inhomogeneities.  In this case, not many line-shape calculations are available, although \citet{dravinsnordlund90a} already discussed some non-LTE effects in a 3D simulation of Procyon.  While line shapes and asymmetries may be very similar, non-LTE effects may lead to different wavelength shifts.  In non-LTE, a decrease of the convective blueshift is caused by a weakening of the most blueshifted line components above the most rapid upflows.  These typically occur in the hottest granules, whose intense ultraviolet flux ionizes the gas above, thus removing the most blueshifted contributions in lines from neutral atomic species.  The more complex geometry in the corrugated surface structures closer to the limb may add additional effects. 

Non-LTE formation of Fe lines in 3D solar granulation are discussed by \citet{lindetal17} and \citet{smithaetal20}, while further effects from including 3D (also horizontal) radiative transfer and magnetic fields were treated by \citet{holzreutersolanki12, holzreutersolanki15} and \citet{ smithaetal21}.  Numerous authors have discussed non-LTE effects in abundance determinations for various atomic species, not least for metal-poor stars, where the modeling of spectral line response in their rather transparent atmospheres comprises a nontrivial problem.  Although it is possible to anticipate substantial insights, non-LTE calculations for complete spectra embracing all atomic species with their humongous number of overlapping energy levels are currently not feasible.  The limits are set not only by computational realities but not least by the incompleteness of laboratory data.  For a general review of non-LTE modeling in 3D atmospheres, see \citet{bergemann14}.

\subsubsection{Dependence on excitation potential}

Among lines of similar strength, a secondary dependence is that on the lower excitation potential of the atomic transition.  High-excitation lines can be expected to preferentially form in volumes of hotter gas.  This does not apply to the very hottest gases, however, where the gas is largely ionized and lines from neutral species thus weakened.  In Fig.\ \ref{fig:excitation_potential}, \ion{Fe}{i} lines of comparable strength are selected from five different models, including those with a very low ($\chi$$\sim$1 eV) and high ($>$\,4 eV) excitation potential, all within the same wavelength region.  To avoid cluttering the plot, line profiles and bisectors are shown for the three stellar disk positions of $\mu$ = 1, 0.41 and 0.09.  The amplitude of bisector shifts increases for the high-excitation line, although the $\chi$-dependence is pronounced only in the somewhat cooler models, where the conditions are somewhat marginal for high-level excitations.  In the hotter stars, the temperatures seemingly are high enough for different excitation levels in neutral species not to make much of a difference, and greater effects might instead be expected between lines in different ionization states. 

As discussed above, there is a further dependence on wavelength region (Fig.\ \ref{fig:color_dependence}), as expected from the granulation contrast being greater in the blue than in the red, although this simple picture may be confounded by specific wavelength-dependent opacities.  Our present line sampling does not suffice to subdivide the lines into separate wavelength bins for each strength and excitation potential; however, the great number of measurable lines in spectra of cool stars should also be adequate for such finer partitioning.

\section{Outlook and outstanding issues}

Present spectra originate from modeling stellar granulation produced by thermally driven convection, simulated in volumes that comprise tiny fractions of any star.  Such granulation can be expected to cover most of the surface of solar-type stars and provides the dominant fraction of stellar flux.  Still, various outstanding issues remain.

Stellar magnetic activity offers both opportunities for its study and challenges in segregating magnetic effects from ordinary surface structure.  As told already in Paper~I, one of the ambitions of the current project is to also recover spatially resolved spectra of starspots and stellar magnetic regions (including their Zeeman-effect signatures in magnetically sensitive lines).  The presence of magnetic flux alters the structure of granulation while greater flux concentrations lead to the formation of starspots, thus contributing spectral lines of different shapes and shifts.  Such effects are further discussed in Paper~V.

Features not yet modeled in hydrodynamic simulations include solar-type supergranulation, that is, structures about a thousand times greater in area than ordinary granules.  Despite only feeble temperature contrasts on the solar surface, their horizontal velocity patterns may well affect line shapes and shifts toward stellar limbs.  The origin of supergranulation has long been the topic of discussion; some recent thoughts are by \citet{cossetterast16}, \citet{cosetteetal17}, \citet{featherstonehindman16}, \citet{rinconetal17} and \citet{viewegetal20}.  Of course, the character of supergranulation on other stars is not known.  Meridional flows may be another two orders of magnitude larger, covering a significant fraction of the star \citep{makarov10, meunierlagrange20a}.  Convection also induces p-mode oscillations and possibly other wave motions, which can modify local irradiance levels and spectral line shapes \citep{zhouetal19}. 

Even given the limitations in current modeling, this survey of realistically complex synthetic spectra across stellar surfaces for stars both cooler and hotter than the Sun, provides quantitative illustrations of how the details of the atmospheric fine structure and granulation dynamics are imprinted onto the detailed spectral line shapes and shifts.  Stellar temperatures slightly hotter or cooler than solar already lead to significant changes in spectral line behavior.  Hotter F-type stars display distinct signatures all over their disks, while the lines in the cooler K-types are only slightly affected (Figs.\ \ref{fig:line_strength}-\ref{fig:excitation_potential}).  The noise-free hyper-high spectral resolution used in this survey preserves the full information of line asymmetries and wavelength shifts, defining optimum targets for future observations.  Salient features of line profiles include how their widths, depths, and wavelength positions change across stellar surfaces.  As we will explore in Paper~V \citep{dravinsetal21}, these are the parameters that are least challenging to observe, also remaining distinguishable in noisier data at the lower resolutions of current spectrometers.

\begin{acknowledgements}
{The work by DD is supported by grants from The Royal Physiographic Society of Lund.  HGL gratefully acknowledges financial support by the Deutsche Forschungsgemeinschaft (DFG, German Research Foundation) -- Project ID~138713538--SFB~881 (`The Milky Way System', subproject A04).  This work has made use of the VALD database, operated at Uppsala University, the Institute of Astronomy RAS in Moscow, and the University of Vienna.  We made use of NASA’s ADS Bibliographic Services and the arXiv$^{\circledR}$ distribution service. We thank the referee for several valuable and insightful comments. }

\end{acknowledgements}


\begin{appendix}

\section{List of hydrodynamic models}

The models used for the hydrodynamic simulations and the ensuing spectral line synthesis are characterized by stellar parameters such as temperature and surface gravity, and by their spatial extent and computational step sizes.  In the text these models were mainly referred to by their approximate spectral type.  The listing of model identifiers in Table \ref{table:co5bold_models} enables their unique identification or comparison with other work based on the CO\,$^5$BOLD grid \citep{freytagetal12}. The solar model (used in Paper~V) is also included.

The number of sampling points for each spectral line profile is the product of horizontal spatial resolution points in $x$ and $y$, the number of temporal snapshots extracted during the simulation sequence, and the number of different directions (combined elevation and azimuth angles) for the emerging radiation.  The $x$-$y$ horizontal ($hsize$) and $z$ vertical ($vsize$) extents of the hydrodynamic simulation volume are here given in centimeter.  

The direction cosines $\mu$ = cos\,$\theta$ for the intensities are computed corresponding to the abscissa of a Gauss-Radau integration scheme. To obtain the flux emitted by the full stellar disk, a weighted sum is computed over the temporally and azimuthally averaged intensities. The weights for each $\mu$ are given by the weights of this Gauss-Radau scheme.  Further computational details are given by \citet{ludwigetal21}.

\begin{table*}
\caption{CO\,$^5$BOLD model identifiers.  The hydrodynamic models and their ensuing spectra 
(Table \ref{table:synthetic_spectra}) have unique identifiers also used in other work.  The spectra were sampled in 47\,$\times$\,47 grids across the simulation areas of 140\,$\times$\,140 horizontal spatial points extending over $hsize$ in $x$ and $y$, at about 20 temporal snapshots, and for up to 21 angular directions.  
}             
\label{table:co5bold_models}      
\centering          
\begin{tabular}{c c c c c c c}
\hline\hline      
\noalign{\smallskip} 
T$_{\textrm{eff}}$ [K] & log~$\varg$ [cgs] & Spectrum \ & Model & Spectral points & boxsize:$hsize$ [cm] & boxsize:$vsize$ [cm] \\ 
\noalign{\smallskip}
\hline  
\noalign{\smallskip}
3964 & 4.5 & hyd0007 -- `K8~V' & d3t40g45mm00n01 & 47 $\times$ 47  $\times$ 19  $\times$ 21 & 4.74 10$^{8}$ & 1.23 10$^{8}$ \\
4982 & 4,5 & hyd0028 -- `K2 V'  & d3t50g45mm00n04 & 47 $\times$  47  $\times$  20  $\times$ 21 & 4.94 10$^{8}$  & 2.48 10$^{8}$ \\
5700 & 4.4 & `Sun' -- `G2~V'    & d3gt57g44n58  & 47 $\times$  47  $\times$ 19 $\times$ 13 & 5.60 10$^{8}$ &  2.25 10$^{8}$ \\
5865 & 4.5 &hyd0052 -- `G1~V' &  d3t59g45mm00n01 &  47 $\times$  47  $\times$  19  $\times$ 21  & 6.02 10$^{8}$  & 3.84 10$^{8}$ \\
6233 & 4.5 & hyd0061 -- `F7~V' &  d3t63g45mm00n01 &  47 $\times$  47  $\times$ 20 $\times$  21 &  7.00 10$^{8}$ & 3.95 10$^{8}$ \\               
6726 & 4.25 & hyd0073 -- `F3~V' & d3t68g43mm00n01 &  47 $\times$  47  $\times$ 20 $\times$  21 & 1.64 10$^{9}$ & 2.42 10$^{9}$ \\  
\noalign{\smallskip}
\hline                  
\end{tabular}
\end{table*}

\section{Listing of selected lines}

List of the 12 groups of selected \ion{Fe}{i} and \ion{Fe}{ii} lines studied, first chosen among essentially unblended lines in the solar spectrum, and then verified to be possible to trace through most stellar spectra.  Parameters are from the VALD database \citep{ryabchikovaetal15} and \citet{stenflolindegren77}.  Line absorption depths are solar disk-center values in units of the normalized local continuum but only signify the line-strength group -- precise line strengths differ among the various stellar models and across stellar disks.

\begin{table}
\tiny
\caption{Selected \ion{Fe}{i} and \ion{Fe}{ii} lines}             
\label{table:table_fe_lines}      
\centering          
\begin{tabular}{c c c c c}
\hline\hline     
\noalign{\smallskip}  
 & ${\lambda}$ [nm] & Solar depth [\%] & ${\chi}$ [eV] & Land{\'e} $\varg$$_{\textrm{eff}}$  \\ 
 \noalign{\smallskip}  
\hline
\\
\multicolumn{5}{l}{ Very strong \ion{Fe}{i}, low ${\chi}$, 500-550 nm } \\
\noalign{\smallskip}
\ion{Fe}{i}  &  507.97462	&  82.1  & 0.99  & 1.5 \\
\ion{Fe}{i}  &  512.73655  &  81.9  & 0.91  &  1.5\\
\ion{Fe}{i}  &  550.14715	&  80.9	& 0.96  &  1.875 \\
\ion{Fe}{i}  &  550.67864  &  82.1  &  0.99  &  2.0 \\ 
\noalign{\smallskip}
\multicolumn{5}{l}{ Very strong \ion{Fe}{i}, high ${\chi}$, 500-550 nm } \\
\noalign{\smallskip}
\ion{Fe}{i}  &  507.47556	&  79.8	& 4.22  &  0.90  \\
\ion{Fe}{i}  &  536.74755  &  80.3  &  4.41  & 0.875 \\
\ion{Fe}{i}  &  538.33792  &  83.3  & 4 .31  & 1.083  \\
\ion{Fe}{i}  &  541.52108	&  82.1  & 4.39  &  1.083  \\
\noalign{\smallskip}
\multicolumn{5}{l}{ Strong \ion{Fe}{i}, high ${\chi}$, 500-550 nm } \\
\noalign{\smallskip}
\ion{Fe}{i}  &  506.71569	 &  67.5	&  4.22  & 1.25 \\
\ion{Fe}{i}  &  519.6065	 &  69.4 	 & 4.26 &  0.667 \\
\ion{Fe}{i}  &  524.37823	 &  62.1	&  4.26 &  1.50 \\
\ion{Fe}{i}  &  539.82860	 &  67.4  & 4.44 &  0.733 \\
\noalign{\smallskip}
\multicolumn{5}{l}{ Weak \ion{Fe}{i}, low ${\chi}$, 465-610 nm } \\
\noalign{\smallskip}
\ion{Fe}{i}  &  467.28364	& 38.1  & 1.61  & 0.25 \\
\ion{Fe}{i}  &  479.87776	 &  39.6  & 1.61 &  1.083 \\
\ion{Fe}{i}  &  512.76876	&  25.2	& 0.05  & 1.00  \\
\ion{Fe}{i}  &  608.27147  &  36.3  & 2.22  & 2.00 \\
\noalign{\smallskip}
\multicolumn{5}{l}{ Weak \ion{Fe}{i}, high ${\chi}$, 500-550 nm } \\
\noalign{\smallskip}
\ion{Fe}{i}  &  501.64778  &  37.4  & 4.26  &  1.50  \\
\ion{Fe}{i}  &  529.5316	&	32.6  & 4.41  &  0.708 \\
\ion{Fe}{i}  &  538.63345	 & 35.7	& 4.15 & 1.167 \\
\ion{Fe}{i}  &  541.70405	& 36.4	& 4.41  &  0.667 \\
\noalign{\smallskip}
\multicolumn{5}{l}{ Very weak \ion{Fe}{i}, high ${\chi}$, 500-550 nm } \\
\noalign{\smallskip}
\ion{Fe}{i}  &  505.84987	 &	14.9  &	 3.64  &  1.333 \\
\ion{Fe}{i}  &  521.38071	&  9.2  & 3.94  & 1.25 \\
\ion{Fe}{i}  &  535.81168	&  11.1 & 3.30  &  1.167 \\
\ion{Fe}{i}  &  542.2156	& 12.0 &  4.32  &  1.50 \\
\noalign{\smallskip}
\multicolumn{5}{l}{ Very strong \ion{Fe}{ii}, medium ${\chi}$, 450-500 nm } \\
\noalign{\smallskip}
\ion{Fe}{ii}  &  449.14035  & 75.0	&  2.85 &  ... \\
\ion{Fe}{ii}  &  450.82866	  & 80.9	&  2.85 &  ... \\
\ion{Fe}{ii}  &  452.02258	  & 78.5	&  2.81 &  ... \\
\ion{Fe}{ii}  &  492.39299	  & 83.8  & 2.89  	&  ... \\
\noalign{\smallskip}
\multicolumn{5}{l}{ Medium \ion{Fe}{ii}, medium ${\chi}$, 440-550 nm } \\
\noalign{\smallskip}
\ion{Fe}{ii}  &   441.35941  &  40.2  & 2.68  & ... \\
\ion{Fe}{ii}  &   465.69787  &  43.0  & 2.89  & ... \\
\ion{Fe}{ii}  &   467.01723  &  35.1  & 2.58  & ... \\
\ion{Fe}{ii}  &   499.33527  &  42.4  & 2.81  & ... \\
\noalign{\smallskip}
\multicolumn{5}{l}{ Medium \ion{Fe}{ii}, medium ${\chi}$, 620-660 nm } \\
\noalign{\smallskip}
\ion{Fe}{ii}  &  623.83903  & 37.5   &  3.89  & ... \\
\ion{Fe}{ii}  &  641.69282  & 36.1   &  3.89  & ... \\
\ion{Fe}{ii}  &  643.26831  & 36.6   &  2.89 & ... \\
\ion{Fe}{ii}  &  651.60855  & 45.8   &  2.89 & ... \\
\noalign{\smallskip}
\multicolumn{5}{l}{ Weak \ion{Fe}{ii}, medium ${\chi}$, 510-610 nm } \\
\noalign{\smallskip}
\ion{Fe}{ii}  &    510.06563  &  21.8  &  2.81 & ... \\
\ion{Fe}{ii}  &    513.26658  &  26.3  &  2.81  & ... \\
\ion{Fe}{ii}  &    525.69346  &  21.9  &  2.89 & ... \\
\ion{Fe}{ii}  &    608.41061  &  20.2  &  3.20  & ... \\
\noalign{\smallskip}
\multicolumn{5}{l}{ Very weak \ion{Fe}{ii}, medium ${\chi}$, 480-640 nm} \\
\noalign{\smallskip}
\ion{Fe}{ii}  &    483.31919  &  12.7  &  2.66 & ... \\
\ion{Fe}{ii}  &    513.67971  &  14.9  &  2.84  & ... \\
\ion{Fe}{ii}  &    611.33221  &  11.3  &  3.22  & ... \\
\ion{Fe}{ii}  &    623.99431  &  12.2  &  3.89  & ... \\
\noalign{\smallskip}
\multicolumn{5}{l}{ Extremely weak \ion{Fe}{ii}, high ${\chi}$, 630-650 nm } \\
\noalign{\smallskip}
\ion{Fe}{ii}  &   638.37171  &  9.1  &  5.55  & ... \\
\ion{Fe}{ii}  &   638.54441  &  3.8  &  5.55  & ... \\
\ion{Fe}{ii}  &   644.29525  &  3.8  &  5.55  & ... \\
\ion{Fe}{ii}  &   644.64102  &  3.9  &  6.22  & ... \\
\noalign{\smallskip}
\hline  \hline  

\end{tabular}
\end{table}

\end{appendix}

\end{document}